\pdfoutput=1
\documentclass[10pt]{article}
\usepackage[T1]{fontenc}
\usepackage[utf8]{inputenc}
\usepackage{lmodern}
\usepackage{amsmath,amssymb}
\usepackage[margin=0.85in]{geometry}
\usepackage{booktabs,array}
\usepackage{graphicx}
\makeatletter
\def\maxwidth{\ifdim\Gin@nat@width>\linewidth\linewidth\else\Gin@nat@width\fi}
\def\maxheight{\ifdim\Gin@nat@height>\textheight\textheight\else\Gin@nat@height\fi}
\makeatother
\setkeys{Gin}{width=\maxwidth,height=\maxheight,keepaspectratio}
\makeatletter
\def\fps@figure{htbp}
\makeatother
\usepackage[dvipsnames]{xcolor}
\usepackage{fancyvrb}

\DefineVerbatimEnvironment{Highlighting}{Verbatim}{commandchars=\\\{\}}
\newenvironment{Shaded}{}{}

\newcommand{\BuiltInTok}[1]{\textcolor[rgb]{0.00,0.50,0.00}{#1}}

\newcommand{\CommentTok}[1]{\textcolor[rgb]{0.38,0.63,0.69}{\textit{#1}}}

\newcommand{\ControlFlowTok}[1]{\textcolor[rgb]{0.00,0.44,0.13}{\textbf{#1}}}

\newcommand{\DecValTok}[1]{\textcolor[rgb]{0.25,0.63,0.44}{#1}}

\newcommand{\KeywordTok}[1]{\textcolor[rgb]{0.00,0.44,0.13}{\textbf{#1}}}
\newcommand{\NormalTok}[1]{#1}
\newcommand{\OperatorTok}[1]{\textcolor[rgb]{0.40,0.40,0.40}{#1}}

\newcommand{\StringTok}[1]{\textcolor[rgb]{0.25,0.44,0.63}{#1}}

\providecommand{\tightlist}{%
  \setlength{\itemsep}{0pt}\setlength{\parskip}{0pt}}
\usepackage[round]{natbib}
\usepackage{url}
\usepackage[
  pdftitle={Lineage Is Memory: Parent Inheritance and Feedback Disclosure in Short-Budget LLM-Guided Optimization},
  pdfauthor={Shanglin Yang, Zhan Shi},
  colorlinks=true,
  linkcolor=Maroon,
  filecolor=Maroon,
  citecolor=Blue,
  urlcolor=Blue
]{hyperref}

\title{Lineage Is Memory: Parent Inheritance and Feedback Disclosure in Short-Budget LLM-Guided Optimization}
\author{
  \textbf{Shanglin Yang}\\
  Independent Researcher\\
  \texttt{kudoysl@gmail.com}
  \and
  \textbf{Zhan Shi}\\
  Santa Clara University\\
  \texttt{ashi2@scu.edu}
}
\date{}

\begin{document}
\maketitle
\begin{abstract}
Iterative language-model optimizers refine prompts and programs by
evaluating a candidate, showing a proposer model a sample of failing
cases, and keeping revisions that improve performance. Recent work
attributes the limits of this loop to narrow single-round reflection,
introducing full-set error clustering, targeted minibatches, or larger
proposer models so the optimizer sees more failure modes per step. Yet
every existing evaluation tests these mechanisms exclusively inside
lineage-preserving search, where each revision edits a surviving parent,
leaving the contribution of parent inheritance confounded with the
breadth of per-round feedback. Across 1,750 pre-registered runs spanning
nine prompt and program synthesis benchmarks, we separate parent
inheritance from feedback disclosure under matched evaluation budgets.
We find that greedy parent inheritance drives the vast majority of
optimization progress, outperforming seed-reset resampling even when
seed-reset proposals receive a clustered summary of all failures in
every round or use a larger proposer model. While full-set error
clustering and diverse failure selection improve single-step rewrites
from the seed and accelerate the opening rounds of search, sequential
parent inheritance rapidly closes this gap as solved error modes are
retained in the candidate and filtered from subsequent failure lists.
Moreover, decomposing token expenditure in GEPA reveals that proposer
reflection accounts for only a small fraction of total cost whereas
validation and minibatch gating consume nearly the entire budget,
showing that greedy parent inheritance over raw failures captures the
gains of complex reflective loops at a fraction of their compute.
\end{abstract}

\hypertarget{introduction}{%
\section{Introduction}\label{introduction}}

Reflective prompt and program optimizers refine candidate instructions
or code through an iterative propose-and-evaluate loop. In GEPA
\citep{a2025gepa}, the system runs a candidate instruction on a small
minibatch, presents the inputs, outputs, and evaluator feedback to a
proposer model, accepts the revision if it improves minibatch
performance, and selects subsequent parents from a Pareto front. Similar
evolutionary loops underlie OPRO \citep{yang2023large}, EvoPrompt
\citep{guo2023evoprompt}, Promptbreeder
\citep{fernando2023promptbreeder}, FunSearch
\citep{romeraparedes2023mathematical}, and AlphaEvolve
\citep{alexander2025alphaevolve}.

Recent work has focused on widening what the proposer observes during
reflection. \citet{liu2026espo} note via a coupon-collector calculation
that a random minibatch of three examples exposes only a small fraction
of error modes per round, and propose ESPO, which clusters all training
errors into three to seven diagnostic patterns before generating edits.
RPT \citep{bayat2026reflective} provides a tool that aggregates
critiques across the training set, AdaGEPA \citep{chen2026adagepa}
steers minibatch slots toward identified weaknesses, Contrastive
Reflection \citep{koh2026contrastive} pairs failing and passing examples
from decision-tree slices, and CASD \citep{singh2026coding} lets a
coding agent inspect the full trace corpus. In all of these iterative
studies, revisions are tested only inside lineage-preserving search,
where every proposal edits a surviving parent. Without a lineage-off
control that resets each proposal to the initial seed, existing
evaluations cannot tell how much progress comes from wider single-round
reflection and how much comes from accumulating edits along a
parent-child lineage.

To separate these factors, we organize a large language model (LLM)
optimizer around three design choices: lineage \(P\), feedback
disclosure \(F\), and evaluation scheduling \(A\). Lineage \(P\)
determines whether each proposal edits the initial seed
\(P = \text{seed}\), which corresponds to best-of-\(N\) resampling from
the seed's feedback, or edits the current search-best candidate
\(P = \text{parent}\). Disclosure \(F\) determines whether the evaluator
shows the proposer only the scalar score, the first \(k\) failures,
\(k\) structurally diverse failures, a random mixed minibatch of \(k\)
passing and failing examples, or an LLM-clustered summary of all
failures with or without expected answers. Scheduling \(A\) governs
candidate evaluation and stopping. We run a crossed \(P \times F\)
factorial experiment under a fixed full-search-set scheduler with early
stopping, complemented by a system-level comparison against GEPA under
matched maximum task-model call caps. Across 1,750 runs on nine task
families, comprising five constructed rule-discovery tasks, a
rubric-judged customer support task, ScoNe \citep{she2023scone}, a
12-intent subset of Banking77 \citep{casanueva2020efficient}, and
BIG-Bench Hard salient translation error detection (BBH-STE)
\citep{suzgun2022challenging}, we replicate the main contrasts on
\texttt{claude-haiku-5-5} and test a larger proposer model
\texttt{gpt-4.1}.

Our main findings are as follows:

\begin{itemize}
\tightlist
\item
  \textbf{Parent inheritance outweighs single-round disclosure breadth
  and proposer size (Sections 4.1--4.2):} On tasks where rules must be
  recovered from feedback, switching from \(P = \text{seed}\) to
  \(P = \text{parent}\) produces the largest accuracy gain in our study,
  adding +0.238 across all seven primary tasks on \texttt{gpt-4.1-mini}
  and +0.222 across five tasks on \texttt{claude-haiku-5-5}. This
  advantage holds even when seed-reset proposals draw fresh random
  examples each round in \texttt{seed-minibatch} or receive a clustered
  summary of all search-set failures in every round in
  \texttt{seed-clustergold}: showing just four raw failures under parent
  inheritance still beats \texttt{seed-clustergold} by +0.157. Upgrading
  the proposer from \texttt{gpt-4.1-mini} to \texttt{gpt-4.1} on three
  feedback-dependent prompt tasks yields no clear gain in either row,
  while parent inheritance under \texttt{gpt-4.1} adds +0.380.
\item
  \textbf{Fact retention in the prompt explains the Banking77 gap with
  GEPA (Section 4.3):} On the 12-intent Banking77 task, where intent
  names appear only in error feedback, baseline \texttt{parent-first}
  initially scored 0.372 against 0.664 for GEPA. Inspection of the
  search logs showed that early formatting-only edits scored 0.000 and
  stayed anchored to the seed, whereas GEPA's reflection prompt
  explicitly tells the proposer to copy domain facts from feedback into
  the instruction. In a follow-up experiment on Banking77 and BBH-STE,
  appending one sentence with this instruction raised
  \texttt{parent-first} on Banking77 from 0.372 to 0.856 at 43\% of
  GEPA's median token cost, while \texttt{seed-first} with the same
  sentence retained only the 4 intents shown in round 1 and scored
  0.317. On BBH-STE, where all six option labels are already in the
  seed, the same sentence reduced accuracy by -0.061.
\item
  \textbf{Parent inheritance closes the gap with wider per-round
  disclosure over multiple rounds (Sections 4.4--4.6):} When resetting
  to the seed, showing only the first four failures crowds out
  zero-feedback code priors on unshown error categories, and both
  diverse sampling and target-preserving clustering improve accuracy
  over \texttt{first-k} disclosure. Under parent inheritance, full-set
  clustering initially leads in round 2 by +0.173, but as solved
  categories are retained in the parent and filtered from future failure
  lists, \texttt{first-k} disclosure shifts across 46\% to 86\% of the
  search set, catching up by round 4 and finishing at the same level.
  Conversely, random mixed minibatches reduce parent-inheritance
  accuracy by spending feedback slots on passing examples.
\item
  \textbf{GEPA spends most of its token budget on validation and
  minibatch gating (Section 4.7):} Across all seven prompt tasks
  evaluated with GEPA, proposer reflection consumes only 2\% to 15\% of
  total tokens, while full search-set validations and minibatch checks
  consume the remaining 85\% to 98\%.
\end{itemize}

\hypertarget{related-work}{%
\section{Related Work}\label{related-work}}

\hypertarget{llm-guided-program-evolution}{%
\subsection{LLM-Guided Program
Evolution}\label{llm-guided-program-evolution}}

Evolutionary search with language models, as in FunSearch
\citep{romeraparedes2023mathematical}, AlphaEvolve
\citep{alexander2025alphaevolve}, and ImprovEvolve
\citep{kravatskiy2026improvevolve}, maintains an archive of high-scoring
programs and prompts a model to mutate or recombine selected parents.
Performance gains in this literature are typically credited to the
evolutionary loop as a whole. A recent evaluation audit of 27
self-evolving configuration and code optimization systems in
recommendation and ranking \citep{yang2026selfevolvingrecsys} finds that
none runs an equal-budget random or seed-reset search over its edit
space or reports multi-session spread on a sealed test split. In general
program evolution, Evolution or Illusion? \citep{oved2026evolution}
varies how a fixed generation budget is split between initial seeds and
evolutionary generations, showing that strategy rankings can reverse
across budget scales. \citet{oved2026evolution} hold feedback
presentation fixed, do not break down token usage by call type, and
treat best-of-\(N\) resampling only as a reference curve. We complement
these studies by fixing the evaluation budget cap, separating search and
held-out splits across ten seeds per cell, and crossing parent
inheritance with feedback disclosure.

\hypertarget{reflective-prompt-optimization}{%
\subsection{Reflective Prompt
Optimization}\label{reflective-prompt-optimization}}

Early automatic prompt optimizers either backpropagated textual
critiques from failing examples, as in ProTeGi
\citep{pryzant2023automatic}, TextGrad \citep{yuksekgonul2024textgrad},
and graph-level LLM-AutoDiff \citep{yin2025llmautodiff}, accumulated
sampling-based momentum across stochastic gradient steps
\citep{ding2026scaling}, or showed the proposer a score-ordered history
of past instructions, as in OPRO \citep{yang2023large} and EvoPrompt
\citep{guo2023evoprompt}. GEPA \citep{a2025gepa} combined reflection on
a small random minibatch of default size \(b=3\) with Pareto-front
parent selection and minibatch acceptance gating. GEPA 0.0.7 also
includes a prompt instruction asking the reflection model to copy
domain-specific facts from feedback into the revised prompt; Section 4.3
tests this instruction directly. Other recent studies examine when
reflective optimization helps or fails: \citet{lumer2026reflective} show
that per-dimension feedback transfers better than scalar scores under
noisy judges, consistent with findings that optimizing harnesses against
ungrounded proxy judges induces Goodhart overfitting without improving
held-out validity \citep{yang2026ideationgoodhart};
\citet{giummarra2026understanding} find that GEPA improves MATH-500
accuracy but fails to improve on HotpotQA with local reasoning models
while increasing token cost; and \citet{ma2024large} show that prompt
optimizers rely heavily on the proposer's pre-trained task prior,
matching our results on \texttt{dates} and \texttt{scone}, where the
model prior already fits the specification.

\hypertarget{expanding-reflection-scope}{%
\subsection{Expanding Reflection
Scope}\label{expanding-reflection-scope}}

Several recent papers argue that small random minibatches under-sample
the error distribution during reflection. \citet{liu2026espo} formalize
this via a coupon-collector bound and introduce ESPO, which clusters all
training failures into three to seven error patterns in a Diagnose stage
before generating pattern-specific edits in a Diversify stage and
bootstrap-selecting among them in a Stabilize stage. RPT
\citep{bayat2026reflective} gives the optimizer a clustering tool over
full-set critiques; AdaGEPA \citep{chen2026adagepa} swaps minibatch
slots toward identified failure modes; CASD \citep{singh2026coding}
replaces iterative search with a coding agent that reads all traces
before writing a single prompt; and Contrastive Reflection
\citep{koh2026contrastive} pairs failing and passing examples from
decision-tree slices. Every iterative method in this group is evaluated
only inside a lineage-preserving loop, without a seed-reset control or a
zero-feedback baseline. Testing disclosure schemes with and without
parent inheritance shows that single-round error coverage strongly
limits seed-reset optimization, whereas under parent inheritance its
terminal effect is much smaller.

\hypertarget{validation-efficiency-and-example-selection}{%
\subsection{Validation Efficiency and Example
Selection}\label{validation-efficiency-and-example-selection}}

Explore More, Validate Less \citep{zhou2026explore} notes that GEPA
wastes evaluations on minibatches where the parent already scores
perfectly, producing no edit, and proposes trimming full-set
validations. Section 4.7 extends this by decomposing GEPA's token budget
across reflection, minibatch gating, and full validation. Choosing a
small representative set of examples to specify program behavior is also
a classic problem in inductive program synthesis
\citep{pu2017selecting}; our \texttt{diverse} condition adapts
farthest-point input selection to iterative error feedback.

\hypertarget{methodology-and-experimental-design}{%
\section{Methodology and Experimental
Design}\label{methodology-and-experimental-design}}

\hypertarget{the-pfa-optimization-loop}{%
\subsection{The P/F/A Optimization
Loop}\label{the-pfa-optimization-loop}}

We study iterative optimization over a candidate space \(\mathcal{C}\)
consisting of natural language system prompts or Python functions,
implemented within the AdalFlow optimization framework
\citep{yin2025llmautodiff}. Each task defines a search set
\(\mathcal{D}_{\text{search}} = \{(x_i, y^*_i)\}_{i=1}^{N_{\text{search}}}\)
partitioned into \(G\) failure categories
\(\{C^{\text{search}}_1, \dots, C^{\text{search}}_G\}\), and a disjoint
held-out test set
\(\mathcal{D}_{\text{holdout}} = \{(x'_j, y'^*_j)\}_{j=1}^{N_{\text{holdout}}}\)
partitioned into
\(\{C^{\text{holdout}}_1, \dots, C^{\text{holdout}}_G\}\) that is never
accessed during search. An evaluator \(E(c, \mathcal{D})\) runs
candidate \(c \in \mathcal{C}\) on \(\mathcal{D}\), returning binary
instance scores \(s_i(c) \in \{0, 1\}\), accuracy
\(S(c, \mathcal{D}) = \frac{1}{|\mathcal{D}|}\sum_{i} s_i(c)\), and
failing indices \(\mathcal{E}(c) = \{i : s_i(c) = 0\}\).

\textbf{Algorithm 1: Controlled P/F/A Optimization Loop}

\begin{enumerate}
\def\labelenumi{\arabic{enumi}.}
\tightlist
\item
  \textbf{Input:} Seed candidate \(c_0 \in \mathcal{C}\), evaluation
  budget cap \(B\), lineage policy
  \(P \in \{\text{seed}, \text{parent}\}\), disclosure operator \(F\),
  search set \(\mathcal{D}_{\text{search}}\), held-out set
  \(\mathcal{D}_{\text{holdout}}\).
\item
  \textbf{Initialize:} Evaluate \(c_0\) on
  \(\mathcal{D}_{\text{search}}\) and cache
  \(E(c_0, \mathcal{D}_{\text{search}}) = \big(S(c_0, \mathcal{D}_{\text{search}}), \mathcal{E}(c_0)\big)\).
  Set history \(\mathcal{H}_0 = \{c_0\}\) and current best
  \(c^*_0 = c_0\).
\item
  \textbf{For} round \(t = 1, \dots, B - 1\) \textbf{do:}

  \begin{itemize}
  \tightlist
  \item
    \textbf{Stop check (\(A\)):} If
    \(S(c^*_{t-1}, \mathcal{D}_{\text{search}}) = 1.0\), stop early.
  \item
    \textbf{Parent selection (\(P\)):} Set \(p_t = c_0\) if
    \(P = \text{seed}\), or \(p_t = c^*_{t-1}\) if
    \(P = \text{parent}\). Retrieve cached
    \(E(p_t, \mathcal{D}_{\text{search}})\).
  \item
    \textbf{Feedback disclosure (\(F\)):} Build feedback string
    \(f_t = F(p_t, \mathcal{D}_{\text{search}}, k=4)\) from
    \(E(p_t, \mathcal{D}_{\text{search}})\).
  \item
    \textbf{Proposal:} Query proposer \(M_{\text{prop}}\) with
    \texttt{PROPOSE\_TEMPLATE} (Appendix B.1) using the task objective,
    \(p_t\), and \(f_t\) to generate \(c_t\).
  \item
    \textbf{Evaluation (\(A\)):} Evaluate \(c_t\) on
    \(\mathcal{D}_{\text{search}}\), consuming 1 evaluation from cap
    \(B\), and cache \(E(c_t, \mathcal{D}_{\text{search}})\). If
    \(S(c_t, \mathcal{D}_{\text{search}}) > S(c^*_{t-1}, \mathcal{D}_{\text{search}})\),
    set \(c^*_t = c_t\); otherwise keep \(c^*_t = c^*_{t-1}\) under
    strict improvement tie-breaking.
  \end{itemize}
\item
  \textbf{Output:} Evaluate \(c^*\) once on
  \(\mathcal{D}_{\text{holdout}}\) and return held-out accuracy
  \(Y = S(c^*, \mathcal{D}_{\text{holdout}})\).
\end{enumerate}

\textbf{Factor 1: Lineage (\(P\)).} Under \(P = \text{seed}\), every
round sets \(p_t = c_0\) and reuses the cached evaluation of \(c_0\).
Under deterministic disclosure schemes such as \texttt{none},
\texttt{first}, and \texttt{diverse}, both \(p_t\) and \(f_t\) stay
fixed across rounds, so the \(B-1\) proposals are independent samples
from \(M_{\text{prop}}\) at default provider temperature conditioned on
\(c_0\). Under \texttt{minibatch}, \texttt{cluster}, and
\texttt{clustergold}, \(f_t\) varies across rounds while still requiring
a single-edit jump from \(c_0\). Under \(P = \text{parent}\), every
round edits the current search-best candidate \(p_t = c^*_{t-1}\),
reusing its cached evaluation if the previous proposal did not strictly
improve on it. When \(S(c_0, \mathcal{D}_{\text{search}}) = 0.0\), as on
\texttt{labels}, \texttt{routing}, \texttt{replies}, and
\texttt{banking77}, strict-improvement tie-breaking keeps \(p_t = c_0\)
until a proposal scores above zero; this occurs in round 1 on
\texttt{routing} and \texttt{replies} and in round 1 or 2 on
\texttt{labels}, but can take several rounds on \texttt{banking77}
(Section 4.3). Once \(c^*_t\) improves over \(c_0\),
\(P = \text{parent}\) combines three mechanisms relative to
\(P = \text{seed}\): text inheritance, which keeps discovered rules in
\(p_{t+1}\); residual-error exposure shift, which removes solved cases
from \(\mathcal{E}(p_{t+1})\) so new categories enter the \(k=4\)
display window; and multi-step edit decomposition, which allows the
proposer to fix a subset of categories per round. Secondary inspiration
candidates are disabled in all cells.

\textbf{Factor 2: Feedback Disclosure (\(F\)).} Failing examples are
formatted with their input, wrong output, and expected answer or grading
rubric (Appendix B). We compare six operators: - \texttt{none}: Masks
the evaluation of \(p_t\) by setting \texttt{score=None}, omitting
instance scores, and rendering \texttt{\{\{\ feedback\ \}\}} as
\texttt{(no\ feedback)}. Because \(p_t = c_0\) carries no
round-dependent signal under \texttt{none}, \texttt{seed-none} is the
shared zero-feedback baseline. - \texttt{first}: Shows
\(S(p_t, \mathcal{D}_{\text{search}})\), per-instance binary scores, and
the first \(k = 4\) failures of \(p_t\) in dataset order, followed by
the count of remaining failures. In round 1, this exposes 2 categories
on \texttt{labels}, \texttt{dates}, \texttt{replies}, and \texttt{size},
up to 3 on \texttt{duration}, and up to 4 on \texttt{routing},
\texttt{banking77}, and \texttt{bbh\_ste}. - \texttt{diverse}: Selects
\(k = 4\) failures via greedy farthest-point sampling under normalized
Levenshtein distance over a structural input signature defined by
\texttt{\_shape} in \texttt{presets.py}, which maps lowercase letters to
\texttt{a}, uppercase letters to \texttt{A}, and digits to \texttt{9}
while preserving punctuation, starting from the first failure. -
\texttt{minibatch}: Draws \(k = 4\) examples uniformly at random without
replacement from \(\mathcal{D}_{\text{search}}\) in each round using
\texttt{random.Random(1000\ +\ r)} for run seed
\(r \in \{0, \dots, 9\}\), and shows whether \(p_t\) passed or failed
each drawn example alongside its expected answer, while still gating
acceptance on the full search set. - \texttt{cluster}: Passes all
failures \(\mathcal{E}(p_t)\) to one call of the clusterer model, which
matches \(M_{\text{prop}}\) and groups them into 3 to 7 patterns,
outputting one line per pattern with a description, failure count, and
one verbatim representative wrong answer (\texttt{CLUSTER\_PROMPT},
Appendix B.3). Clustering tokens are included in search tokens. -
\texttt{clustergold}: Identical to \texttt{cluster}, except the prompt
requires each representative failure to include the input, wrong answer,
and full expected answer (\texttt{CLUSTER\_GOLD\_PROMPT}, Appendix B.3).

\textbf{Scheduling (\(A\)) and GEPA Comparison.} In our \(P \times F\)
grid, every proposal is evaluated on all of
\(\mathcal{D}_{\text{search}}\) up to \(B = 8\) evaluations for prompt
tasks, corresponding to \(c_0\) plus 7 proposals and at most
\(8 \times N_{\text{search}}\) task calls, and \(B = 12\) evaluations
for code tasks, corresponding to \(c_0\) plus 11 proposals executed with
a 60-second subprocess timeout across 24 test cases for at most 288
executions. Missing fenced blocks trigger at most one retry, though all
1,750 runs produced valid blocks on the first try; code exceptions score
\(0.0\) with the traceback included in feedback. Proposer, task, and
clusterer models use default provider sampling in \texttt{openai} SDK
2.44.0 for \texttt{gpt-4.1-mini} and \texttt{gpt-4.1} and
\texttt{anthropic} SDK 0.112.0 with default adaptive thinking for
\texttt{claude-haiku-5-5}, while the \texttt{replies} judge uses
\texttt{gpt-4.1-mini} at \texttt{temperature=0} with response caching.
For system-level comparison, we run \texttt{gepa} 0.0.7
\citep{a2025gepa} with default minibatch size \(b = 3\), Pareto
selection, minibatch acceptance gating, its built-in reflection prompt,
and the same task-call cap
\(\text{max\_metric\_calls} = 8 \times N_{\text{search}}\), testing both
unmodified \texttt{GEPA\ mb3} and early-stopped
\texttt{GEPA\ mb3\ +\ stop}.

\hypertarget{benchmark-suite-and-task-specifications}{%
\subsection{Benchmark Suite and Task
Specifications}\label{benchmark-suite-and-task-specifications}}

Table 1 summarizes the seven primary tasks and two public extension
tasks.

\begin{center}
\small\setlength{\tabcolsep}{4.5pt}\renewcommand{\arraystretch}{1.18}
\resizebox{\ifdim\width>\textwidth\textwidth\else\width\fi}{!}{%
\begin{tabular}{@{}llcccccc>{\raggedright\arraybackslash}p{5.4cm}@{}}
\toprule
\textbf{Task Family} & \textbf{Artifact Type} & \parbox[b]{1.95cm}{\centering\textbf{Categories (\(G\))}} & \parbox[b]{1.95cm}{\centering\textbf{\(|\mathcal{D}_{\text{search}}|\)}} & \parbox[b]{1.95cm}{\centering\textbf{\(|\mathcal{D}_{\text{holdout}}|\)}} & \parbox[b]{1.95cm}{\centering\textbf{Budget Cap (\(B\))}} & \parbox[b]{1.95cm}{\centering\textbf{Max Task Calls}} & \parbox[b]{1.95cm}{\centering\textbf{Seed Score (Search / Holdout)}} & \textbf{Latent Specification Withheld from \(c_0\) and Revealed via Feedback} \\
\midrule
\texttt{labels} & Prompt & 12 & 24 (2/cat) & 24 (2/cat) & 8 & 192 & 0.00 / 0.00 & Exact \texttt{snake\_case} category names (\texttt{ipv4}, \texttt{hex\_color}, \texttt{us\_date}, \texttt{semver}, etc.) \\
\texttt{dates} & Prompt & 12 & 24 (2/cat) & 24 (2/cat) & 8 & 192 & 0.75 / 0.75 & Slash vs.~dot day/month order, 2-digit century cutoff (19xx vs.~20xx), \texttt{invalid} token \\
\texttt{routing} & Prompt & 6 & 16 (2--3/cat) & 12 (2/cat) & 8 & 128 & 0.00 / 0.00 & 6 queue names (\texttt{payments}, \texttt{fulfillment}, \texttt{logistics}, \texttt{trust\_safety}, \texttt{privacy}, \texttt{care}) \& edge policies \\
\texttt{replies} & Prompt (Free-text) & 8 & 16 (2/cat) & 16 (2/cat) & 8 & 128 & 0.00 / 0.00 & 8 Acme support policies (refund window, \textgreater7-day \$10 credit threshold, URLs, legal email, GDPR rules) \\
\texttt{scone} & Prompt (Public NLI) & 6 & 36 (6/cat) & 72 (12/cat) & 8 & 288 & 0.89 / 0.89 & 6 negation-scope conditions from ScoNe \citep{she2023scone}; binary \texttt{entailment} / \texttt{neutral} labels \\
\texttt{duration} & Code (Python) & 7 & 24 & 24 & 12 & 288 & 0.46 / 0.42 & Plain seconds, compound units (\texttt{1h\ 30m}), English words, clock (\texttt{M:SS}, \texttt{HH:MM:SS}), ISO 8601 (\texttt{PT1H30M}), \texttt{ValueError} \\
\texttt{size} & Code (Python) & 7 & 24 & 24 & 12 & 288 & 0.21 / 0.21 & Plain bytes, SI decimal (\texttt{10KB}), IEC binary (\texttt{4MiB}, \texttt{2Gi}), single-letter (\texttt{4K}), comma separators (\texttt{1,024}), English words \\
\texttt{banking77} & Prompt (Public Cls.) & 12 & 24 (2/cat) & 36 (3/cat) & 8 & 192 & 0.00 / 0.00 & 12 intent labels sampled from Banking77 \citep{casanueva2020efficient}, withheld from the initial seed \\
\texttt{bbh\_ste} & Prompt (Public QA) & 6 & 18 (3/cat) & 36 (6/cat) & 8 & 144 & 0.56 / 0.56 & 6 translation error classes in BBH-STE \citep{suzgun2022challenging}; all 6 option names given in seed \\
\bottomrule
\end{tabular}%
}
\end{center}

\textbf{Table 1:} Specifications of the seven primary tasks and two
public extension tasks.

In the five constructed tasks, namely \texttt{labels}, \texttt{dates},
\texttt{routing}, \texttt{duration}, and \texttt{size}, as well as in
\texttt{replies}, the seed \(c_0\) omits key rules or label names so
that high accuracy requires learning from feedback (Appendix B.2). For
the two code tasks, \(c_0\) is a 4-line function that handles only
single-character suffixes: \texttt{s}, \texttt{m}, and \texttt{h} on
\texttt{duration}, and \texttt{B}, \texttt{K}, \texttt{M}, and
\texttt{G} on \texttt{size}. This setup favors methods that accumulate
rules across rounds. To test public benchmarks, we included
\texttt{scone} \citep{she2023scone} in Batch A.1 and added
\texttt{banking77} \citep{casanueva2020efficient} and \texttt{bbh\_ste}
\citep{suzgun2022challenging} in Batch A.4. For \texttt{banking77}, 12
of 77 intents were sampled with \texttt{random.Random(0)} before
registration after an excluded 20-intent smoke test scored at most 0.15
within 8 evaluations, drawing 2 search and 3 held-out queries per
intent. For \texttt{bbh\_ste}, the six option names were included in
\(c_0\), and examples exceeding the 300-character feedback truncation
limit were filtered before registration, retaining 198 of 250 examples
and sampling 3 search and 6 held-out examples per class.

\hypertarget{statistical-estimands-bootstrap-protocol-and-run-accounting}{%
\subsection{Statistical Estimands, Bootstrap Protocol, and Run
Accounting}\label{statistical-estimands-bootstrap-protocol-and-run-accounting}}

The replication unit is an optimization run with seed
\(r \in \{0, \dots, 9\}\), giving \(n = 10\) runs per cell. Let
\(Y_{\tau,P,F,r} \in [0, 1]\) denote held-out accuracy
\(S(c^*, \mathcal{D}_{\text{holdout}})\) on task
\(\tau \in \mathcal{T}\), reserving \(t\) for optimization rounds, and
let \(\bar{Y}_{\tau,P,F} = \frac{1}{10}\sum_{r=0}^9 Y_{\tau,P,F,r}\)
denote the cell mean. Weighting tasks in \(\mathcal{T}\) equally, we
estimate three quantities:

\begin{enumerate}
\def\labelenumi{\arabic{enumi}.}
\tightlist
\item
  \textbf{Lineage main effect (\(\Delta_{\text{lineage}}\)):}
  \[\Delta_{\text{lineage}} = \frac{1}{|\mathcal{T}| \cdot |\mathcal{F}|} \sum_{\tau \in \mathcal{T}} \sum_{F \in \mathcal{F}} \left(\bar{Y}_{\tau,\text{parent},F} - \bar{Y}_{\tau,\text{seed},F}\right).\]
\item
  \textbf{Attenuation interaction (\(I_X\)):} For
  \(X \in \{\text{diverse}, \text{minibatch}, \text{cluster}, \text{clustergold}\}\)
  and
  \(\delta_{\tau,P}(X) = \bar{Y}_{\tau,P,X} - \bar{Y}_{\tau,P,\text{first}}\),
  \[I_X = \frac{1}{|\mathcal{T}|} \sum_{\tau \in \mathcal{T}} \big(\delta_{\tau,\text{seed}}(X) - \delta_{\tau,\text{parent}}(X)\big).\]
  A positive \(I_X > 0\) means disclosure \(X\) improves over
  \texttt{first} more under seed resetting than under parent
  inheritance.
\item
  \textbf{Unshown-category proposal coverage
  (\(\text{Cov}_{\text{unshown}}\)):} In the \texttt{seed} row,
  \(\mathcal{U}_\tau \subset \{1, \dots, G\}\) is the set of categories
  where \(c_0\) fails at least one search example in \(>50\%\) of
  \texttt{seed} runs, excluding categories shown among the first four
  failures in \(>50\%\) of \texttt{seed-first} runs, with
  \texttt{seed-none} using the same reference set \(\mathcal{U}_\tau\).
  Across proposals \(\mathcal{V}_r = \{c_1, \dots, c_{m_r}\}\) in run
  \(r\),
  \[\text{Cov}_{\text{unshown}}(r) = \frac{1}{|\mathcal{U}_\tau|} \sum_{g \in \mathcal{U}_\tau} \frac{1}{|\mathcal{V}_r|} \sum_{c \in \mathcal{V}_r} \mathbf{1}\!\left[s_i(c) = 1 \text{ for all } i \in C^{\text{eval}}_g\right],\]
  where \(C^{\text{eval}}_g\) is \(C^{\text{search}}_g\) for prompt
  tasks and \(C^{\text{holdout}}_g\), evaluated offline after search
  completes, for code tasks. Crowding-out (\(\mathcal{H}_{C1}\)) holds
  when
  \(\bar{\text{Cov}}_{\text{unshown}}(\text{seed-first}) - \bar{\text{Cov}}_{\text{unshown}}(\text{seed-none}) < 0\).
\end{enumerate}

\textbf{Bootstrap inference and run accounting.} All confidence
intervals are 95\% percentile bootstrap intervals given by the 1-indexed
order statistics \([\tilde{\psi}_{(251)}, \tilde{\psi}_{(9750)}]\) over
\(N_{\text{boot}} = 10{,}000\) task-stratified resamples initialized
with \texttt{random.Random(0)} in \texttt{results/analyze.py},
resampling 10 runs with replacement within each task-by-condition cell.
For conditional tokens-to-solve ratios (\(\mathcal{H}_{C3}\)), solved
runs are resampled within each arm on tasks with at least \(3/10\)
solved runs per arm. Alongside these fixed-strata intervals, we report
per-task intervals (Table 2A), leave-one-task-out intervals (Table 8),
and two-stage task-cluster bootstrap intervals that resample tasks at
the outer level: on \texttt{gpt-4.1-mini}, the 7-task cluster interval
is {[}+0.104, +0.363{]} for \(\Delta_{\text{lineage}}\) and {[}+0.052,
+0.336{]} for \(I_{\text{clustergold}}\), which is positive on all 7
tasks with exact two-sided sign test \(p = 0.0156\). All 1,750 completed
runs in \texttt{results/runs.jsonl} come from five sequential
pre-registered batches documented in Table 7 and Appendix A: Batch A.1
with 750 runs, Batch A.2 with 140 runs, Batch A.3 with 350 runs, Batch
A.4 with 470 runs, and Batch A.5 with 40 runs.

\hypertarget{experimental-results}{%
\section{Experimental Results}\label{experimental-results}}

\begin{figure}
\centering
\includegraphics{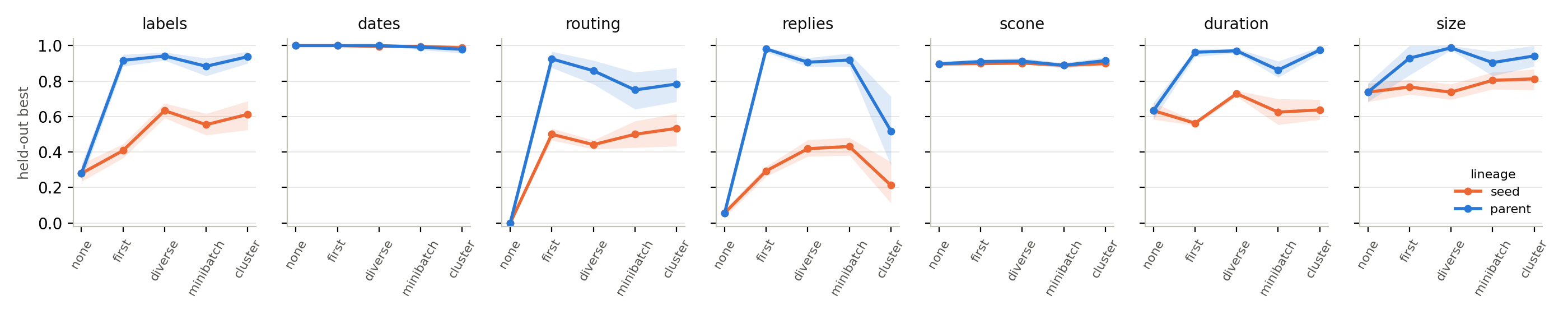}
\caption{Held-out accuracy of the search-best candidate across
disclosure schemes on the seven primary tasks for \texttt{gpt-4.1-mini},
comparing seed-reset resampling (\(P=\text{seed}\), orange) against
greedy parent inheritance (\(P=\text{parent}\), blue). Points show means
over \(n=10\) runs per cell; shaded bands indicate 95\% bootstrap
confidence intervals. The zero-feedback baseline (\texttt{none}) is
shared by both rows.}
\end{figure}

\hypertarget{parent-inheritance-across-task-families-and-models}{%
\subsection{Parent Inheritance Across Task Families and
Models}\label{parent-inheritance-across-task-families-and-models}}

Tables 2A and 2B report held-out accuracy across the seven primary tasks
on \texttt{gpt-4.1-mini}.

\begin{center}
\small\setlength{\tabcolsep}{4.5pt}\renewcommand{\arraystretch}{1.18}
\resizebox{\ifdim\width>\textwidth\textwidth\else\width\fi}{!}{%
\begin{tabular}{@{}>{\raggedright\arraybackslash}p{3.1cm}ccccccccc@{}}
\toprule
\textbf{Lineage (\(P\)) \& Disclosure (\(F\))} & \textbf{\texttt{labels}} & \textbf{\texttt{dates}} & \textbf{\texttt{routing}} & \textbf{\texttt{replies}} & \textbf{\texttt{scone}} & \textbf{\texttt{duration}} & \textbf{\texttt{size}} & \textbf{Mean (All 7)} & \parbox[b]{1.95cm}{\centering\textbf{Mean (5 Non-Ceil.)}} \\
\midrule
\texttt{seed-none} (Zero feedback) & 0.279 & 1.000 & 0.000 & 0.056 & 0.897 & 0.633 & 0.738 & 0.515 & 0.341 \\
\texttt{seed-first} & 0.408 & 1.000 & 0.500 & 0.294 & 0.899 & 0.562 & 0.767 & 0.633 & 0.506 \\
\texttt{seed-diverse} & 0.633 & 0.996 & 0.442 & 0.419 & 0.901 & 0.729 & 0.738 & 0.694 & 0.592 \\
\texttt{seed-minibatch} & 0.554 & 0.996 & 0.500 & 0.431 & 0.889 & 0.625 & 0.804 & 0.686 & 0.583 \\
\texttt{seed-cluster} & 0.613 & 0.988 & 0.533 & 0.212 & 0.899 & 0.637 & 0.812 & 0.671 & 0.561 \\
\texttt{seed-clustergold} (Batch A.2) & 0.683 & 1.000 & 0.625 & 0.794 & 0.908 & 0.637 & 0.875 & 0.789 & 0.723 \\
\texttt{parent-first} & 0.917 & 1.000 & 0.925 & 0.981 & 0.910 & 0.963 & 0.929 & 0.946 & 0.943 \\
\texttt{parent-diverse} & 0.942 & 1.000 & 0.858 & 0.906 & 0.912 & 0.971 & 0.988 & 0.940 & 0.933 \\
\texttt{parent-minibatch} & 0.883 & 0.992 & 0.750 & 0.919 & 0.889 & 0.863 & 0.904 & 0.886 & 0.864 \\
\texttt{parent-cluster} & 0.938 & 0.979 & 0.783 & 0.519 & 0.915 & 0.975 & 0.942 & 0.864 & 0.831 \\
\texttt{parent-clustergold} (Batch A.2) & 0.958 & 0.996 & 0.842 & 0.869 & 0.900 & 0.971 & 0.963 & 0.928 & 0.921 \\
\(\Delta_{\text{lineage}}\) (\texttt{first}, \texttt{diverse}, \texttt{minibatch}, \texttt{cluster}) & \textbf{+0.368} & -0.002 & \textbf{+0.335} & \textbf{+0.492} & +0.010 & \textbf{+0.304} & \textbf{+0.160} & \textbf{+0.238} & \textbf{+0.332} \\
95\% Bootstrap CI for \(\Delta_{\text{lineage}}\) & {[}+0.334, +0.402{]} & {[}-0.009, +0.004{]} & {[}+0.283, +0.387{]} & {[}+0.428, +0.552{]} & {[}+0.000, +0.020{]} & {[}+0.277, +0.330{]} & {[}+0.121, +0.199{]} & {[}+0.224, +0.252{]} & {[}+0.312, +0.351{]} \\
\bottomrule
\end{tabular}%
}
\end{center}

\textbf{Table 2A:} Held-out accuracy cell means and per-task lineage
effect \(\Delta_{\text{lineage}}\) on \texttt{gpt-4.1-mini} (\(n=10\)
runs per cell). ``5 Non-Ceil.'' averages over the five tasks with
headroom above \texttt{seed-none} (\texttt{labels}, \texttt{routing},
\texttt{replies}, \texttt{duration}, \texttt{size}), excluding
\texttt{dates} and \texttt{scone}.

\begin{center}
\small\setlength{\tabcolsep}{4.5pt}\renewcommand{\arraystretch}{1.18}
\resizebox{\ifdim\width>\textwidth\textwidth\else\width\fi}{!}{%
\begin{tabular}{@{}>{\raggedright\arraybackslash}p{8.8cm}cc@{}}
\toprule
Cross-Task Contrast (7 Tasks Equal-Weighted, \texttt{gpt-4.1-mini}) & Point Estimate \& 95\% Bootstrap CI & 5 Non-Ceiling Tasks Mean \\
\midrule
Lineage main effect \(\Delta_{\text{lineage}}\) (\texttt{first}, \texttt{diverse}, \texttt{minibatch}, \texttt{cluster}) & \textbf{+0.238} {[}+0.224, +0.252{]} & \textbf{+0.332} {[}+0.312, +0.351{]} \\
Lineage main effect across all 5 schemes / across \texttt{clustergold} only & +0.218 {[}+0.204, +0.233{]} / +0.139 {[}+0.096, +0.181{]} & +0.305 / +0.198 \\
Cross-row contrast: \texttt{parent-first} \(-\) \texttt{seed-clustergold} / \texttt{parent-first} \(-\) \texttt{seed-minibatch} & \textbf{+0.157} {[}+0.124, +0.195{]} / \textbf{+0.261} {[}+0.235, +0.285{]} & +0.220 / +0.360 \\
Attenuation interaction \(I_{\text{diverse}} = (\text{diverse} - \text{first})_{\text{seed}} - (\text{diverse} - \text{first})_{\text{parent}}\) (\(\mathcal{H}_{C2}\)) & \textbf{+0.068} {[}+0.043, +0.092{]} & +0.096 \\
Attenuation interaction \(I_{\text{minibatch}} = (\text{minibatch} - \text{first})_{\text{seed}} - (\text{minibatch} - \text{first})_{\text{parent}}\) & \textbf{+0.113} {[}+0.079, +0.148{]} & +0.156 \\
Attenuation interaction \(I_{\text{cluster}} = (\text{cluster} - \text{first})_{\text{seed}} - (\text{cluster} - \text{first})_{\text{parent}}\) & \textbf{+0.120} {[}+0.073, +0.167{]} & +0.167 \\
Attenuation interaction \(I_{\text{clustergold}} = (\text{clustergold} - \text{first})_{\text{seed}} - (\text{clustergold} - \text{first})_{\text{parent}}\) & \textbf{+0.174} {[}+0.127, +0.222{]} & +0.239 \\
Seed row within-row contrasts: \texttt{seed-diverse} \(-\) \texttt{seed-first} / \texttt{seed-clustergold} \(-\) \texttt{seed-first} & +0.061 {[}+0.046, +0.077{]} / +0.156 {[}+0.122, +0.188{]} & +0.086 / +0.217 \\
Seed row within-row contrasts: \texttt{seed-minibatch} \(-\) \texttt{seed-first} / \texttt{seed-cluster} \(-\) \texttt{seed-first} & +0.053 {[}+0.030, +0.075{]} / +0.038 {[}+0.009, +0.067{]} & +0.077 / +0.055 \\
Parent row within-row contrasts: \texttt{parent-diverse} \(-\) \texttt{parent-first} / \texttt{parent-clustergold} \(-\) \texttt{parent-first} (\(\mathcal{H}_{S2\text{-gold}}\)) & -0.007 {[}-0.025, +0.012{]} / -0.018 {[}-0.053, +0.010{]} & -0.010 / -0.022 \\
Parent row within-row contrasts: \texttt{parent-minibatch} \(-\) \texttt{parent-first} / \texttt{parent-cluster} \(-\) \texttt{parent-first} & -0.061 {[}-0.086, -0.035{]} / -0.082 {[}-0.119, -0.045{]} & -0.079 / -0.112 \\
\bottomrule
\end{tabular}%
}
\end{center}

\textbf{Table 2B:} Cross-task interaction and within-row disclosure
contrasts on \texttt{gpt-4.1-mini}.

Parent inheritance produces the largest accuracy contrast in the grid,
increasing 7-task held-out accuracy by
\(\Delta_{\text{lineage}} = +0.238\) across the four Batch A.1
disclosure schemes and by \(+0.218\) across all five schemes (Table 2B).
Importantly, this gain is not simply a consequence of
\texttt{seed-first} observing the same four failures each round. Even
when seed-reset proposals draw fresh random examples every round in
\texttt{seed-minibatch} or receive a target-preserving summary of all
search-set failures in every round in \texttt{seed-clustergold},
\texttt{parent-first} beats both by +0.261 and +0.157, and
\texttt{parent-clustergold} beats \texttt{seed-clustergold} by +0.139
across all seven tasks and +0.198 across the five feedback-dependent
tasks. Thus, even when the proposer sees a summary of all failing
categories, rewriting the seed from scratch in a single step falls well
short of accumulating edits across rounds.

Across tasks, Figure 1 and Table 2A show two distinct regimes. On
\texttt{dates} and \texttt{scone}, the model's pre-trained prior already
solves most cases without feedback, scoring 1.000 and 0.897 under
\texttt{seed-none} and leaving little room for lineage to help. On the
five tasks where rules or policies are withheld from \(c_0\), parent
inheritance adds between +0.160 on \texttt{size} and +0.492 on
\texttt{replies}, averaging +0.332. Every pre-registered \texttt{parent}
cell on these five tasks beats every pre-registered \texttt{seed} cell;
the only crossover involves \texttt{seed-clustergold} beating
\texttt{parent-cluster} on \texttt{replies}, where the unadjusted
\texttt{cluster} prompt dropped grading rubrics (Section 4.6).
Replicating the experiment on \texttt{claude-haiku-5-5} across the same
five tasks confirms these patterns (Table 3, Panel A): parent
inheritance adds +0.222 overall and +0.257 when excluding
\texttt{duration}, where Haiku starts near ceiling at 0.954, and
\texttt{parent-first} again outperforms both \texttt{seed-minibatch} and
\texttt{seed-clustergold}.

\begin{center}
\small\setlength{\tabcolsep}{4.5pt}\renewcommand{\arraystretch}{1.18}
\resizebox{\ifdim\width>\textwidth\textwidth\else\width\fi}{!}{%
\begin{tabular}{@{}lccccccc@{}}
\toprule
\parbox[b]{2.85cm}{\raggedright\textbf{Model Configuration \& Cell}} & \textbf{\texttt{labels}} & \textbf{\texttt{routing}} & \textbf{\texttt{replies}} & \textbf{\texttt{duration}} & \textbf{\texttt{size}} & \parbox[b]{2.25cm}{\centering\textbf{Mean (5 Tasks)}} & \parbox[b]{2.25cm}{\centering\textbf{Mean (3 Prompt Tasks)}} \\
\midrule
\multicolumn{8}{@{}l@{}}{\textbf{Panel A: \texttt{claude-haiku-5-5} Replication (\(n=10\) runs per cell, 450 runs across Batches A.3 \& A.4)}} \\
\midrule
\texttt{seed-none} & 0.671 & 0.000 & 0.044 & 0.954 & 0.650 & 0.464 & 0.238 \\
\texttt{seed-first} & 0.725 & 0.500 & 0.312 & 0.863 & 0.817 & 0.643 & 0.512 \\
\texttt{seed-diverse} (Batch A.4) & 0.871 & 0.400 & 0.487 & 0.942 & 0.879 & 0.716 & 0.586 \\
\texttt{seed-minibatch} & 0.775 & 0.508 & 0.444 & 0.942 & 0.863 & 0.706 & 0.576 \\
\texttt{seed-clustergold} & 0.904 & 0.792 & 0.637 & 0.942 & 0.838 & 0.823 & 0.778 \\
\texttt{parent-first} & 0.958 & 0.950 & 0.988 & 1.000 & 1.000 & 0.979 & 0.965 \\
\texttt{parent-diverse} (Batch A.4) & 0.925 & 0.942 & 0.781 & 1.000 & 1.000 & 0.930 & 0.883 \\
\texttt{parent-minibatch} & 0.950 & 0.758 & 0.762 & 0.988 & 0.971 & 0.886 & 0.823 \\
\texttt{parent-clustergold} & 0.992 & 0.917 & 0.950 & 1.000 & 1.000 & 0.972 & 0.953 \\
Haiku \(\mathcal{H}_{H1}\) (\(\Delta_{\text{lineage}}\)) & +0.165 & +0.275 & +0.435 & +0.081 & +0.151 & \textbf{+0.222} {[}+0.204, +0.239{]} & +0.292 \\
Haiku \(\mathcal{H}_{H2}\) (\(I_{\text{clustergold}}\)) & +0.146 & +0.325 & +0.363 & +0.079 & +0.021 & \textbf{+0.187} {[}+0.152, +0.221{]} & +0.278 \\
Haiku \(\mathcal{H}_{H3}\) (\texttt{parent-cgold} \(-\) \texttt{first}) & +0.033 & -0.033 & -0.038 & +0.000 & +0.000 & \textbf{-0.008} {[}-0.017, +0.002{]} & -0.013 \\
Haiku \(\mathcal{H}_{Hd}\) (\(I_{\text{diverse}}\)) & +0.179 & -0.092 & +0.381 & +0.079 & +0.062 & \textbf{+0.122} {[}+0.088, +0.158{]} & +0.156 \\
\midrule
\multicolumn{8}{@{}l@{}}{\textbf{Panel B: Stronger Proposer (\texttt{gpt-4.1} Proposer \& Clusterer, \texttt{gpt-4.1-mini} Task Model, \(n=10\), 150 runs)}} \\
\midrule
\texttt{seed-none} (\texttt{gpt-4.1} prop.) & 0.279 & 0.000 & 0.062 & -- & -- & -- & 0.114 (\(\Delta_{\text{prop}}\): +0.002) \\
\texttt{seed-first} (\texttt{gpt-4.1} prop.) & 0.383 & 0.500 & 0.300 & -- & -- & -- & 0.394 (\(\Delta_{\text{prop}}\): -0.006) \\
\texttt{seed-clustergold} (\texttt{gpt-4.1} prop.) & 0.596 & 0.617 & 0.681 & -- & -- & -- & 0.631 (\(\Delta_{\text{prop}}\): -0.069) \\
\texttt{parent-first} (\texttt{gpt-4.1} prop.) & 0.904 & 0.892 & 0.969 & -- & -- & -- & 0.922 (\(\Delta_{\text{prop}}\): -0.019) \\
\texttt{parent-clustergold} (\texttt{gpt-4.1} prop.) & 0.858 & 0.792 & 0.944 & -- & -- & -- & 0.865 (\(\Delta_{\text{prop}}\): -0.025) \\
Strong-Prop. \(\mathcal{H}_{S1\text{-prop}}\) (\(\Delta_{\text{lineage}}\)) & +0.392 & +0.283 & +0.466 & -- & -- & -- & \textbf{+0.380} {[}+0.337, +0.425{]} \\
Strong-Prop. \(\mathcal{H}_{S2\text{-prop}}\) (\(I_{\text{clustergold}}\)) & +0.258 & +0.217 & +0.406 & -- & -- & -- & \textbf{+0.294} {[}+0.201, +0.381{]} \\
Strong-Prop. \(\mathcal{H}_{S3\text{-prop}}\) (\texttt{parent-cgold} \(-\) \texttt{first}) & -0.046 & -0.100 & -0.025 & -- & -- & -- & \textbf{-0.057} {[}-0.101, -0.016{]} \\
\bottomrule
\end{tabular}%
}
\end{center}

\textbf{Table 3:} Cross-model replication on \texttt{claude-haiku-5-5}
(Panel A) and stronger-proposer evaluation using \texttt{gpt-4.1} as
proposer and clusterer with \texttt{gpt-4.1-mini} as task model (Panel
B). Here \(\Delta_{\text{prop}}\) is the difference
(\(\text{gpt-4.1 proposer} - \text{gpt-4.1-mini proposer}\)) across
\texttt{labels}, \texttt{routing}, and \texttt{replies}, and
\(\mathcal{H}_{S3\text{-prop}}\) is \texttt{parent-clustergold} \(-\)
\texttt{parent-first}.

\hypertarget{upgrading-proposer-model-capacity-versus-parent-inheritance}{%
\subsection{Upgrading Proposer Model Capacity Versus Parent
Inheritance}\label{upgrading-proposer-model-capacity-versus-parent-inheritance}}

Panel B of Table 3 tests whether upgrading the proposer and clusterer
from \texttt{gpt-4.1-mini} to \texttt{gpt-4.1} reduces the need for
parent inheritance on \texttt{labels}, \texttt{routing}, and
\texttt{replies} while holding the task model fixed at
\texttt{gpt-4.1-mini}. Across all five tested conditions, replacing
\texttt{gpt-4.1-mini} with \texttt{gpt-4.1} yields no positive accuracy
gain, with \(\Delta_{\text{prop}}\) ranging from -0.069 in
\texttt{seed-clustergold} to +0.002 in \texttt{seed-none}. On
\texttt{labels} under \texttt{seed-none}, both proposers average
\(67/240 = 0.279\) from distinct run distributions, where nine of ten
\texttt{gpt-4.1} runs score \(6/24 = 0.250\) and one scores
\(13/24 = 0.542\). Meanwhile, parent inheritance under the
\texttt{gpt-4.1} proposer adds \(\Delta_{\text{lineage}} = +0.380\),
exhibits a large attenuation interaction of
\(I_{\text{clustergold}} = +0.294\), and scores slightly higher with
\texttt{first-k} failures than with \texttt{clustergold} by +0.057.
Upgrading proposer capacity within the GPT-4.1 family therefore does not
substitute for parent inheritance.

\hypertarget{public-benchmarks-and-instruction-level-retention-of-feedback-facts}{%
\subsection{Public Benchmarks and Instruction-Level Retention of
Feedback
Facts}\label{public-benchmarks-and-instruction-level-retention-of-feedback-facts}}

Table 4 reports the 9-cell factorial grid alongside GEPA on the
12-intent Banking77 subset and BBH-STE.

\begin{center}
\small\setlength{\tabcolsep}{4.5pt}\renewcommand{\arraystretch}{1.18}
\resizebox{\ifdim\width>\textwidth\textwidth\else\width\fi}{!}{%
\begin{tabular}{@{}lcccccc@{}}
\toprule
\parbox[b]{2.85cm}{\raggedright\textbf{Condition / Method}} & \parbox[b]{2.25cm}{\centering\textbf{\texttt{banking77} Held-Out}} & \parbox[b]{2.25cm}{\centering\textbf{\texttt{bbh\_ste} Held-Out}} & \textbf{Pooled Mean} & \parbox[b]{2.25cm}{\centering\textbf{\texttt{banking77} Intents Named in Best Prompt (Run Distribution; Mean / 12)}} & \parbox[b]{2.25cm}{\centering\textbf{\texttt{banking77} Search Tokens / Run (Median / Mean)}} & \parbox[b]{2.25cm}{\centering\textbf{\texttt{bbh\_ste} Search Tokens / Run (Median / Mean)}} \\
\midrule
\multicolumn{7}{@{}l@{}}{\textbf{Panel A: Pre-Registered Public Benchmark Grid (\texttt{gpt-4.1-mini}, \(n=10\) runs per cell, Batch A.4)}} \\
\midrule
\texttt{seed-none} & 0.000 & 0.556 & 0.278 & 0.0 / 12 & 21.7k / 22.0k & 42.7k / 42.8k \\
\texttt{seed-first} & 0.225 & 0.631 & 0.428 & 3.2 / 12 (range 1--4) & 25.5k / 26.0k & 48.2k / 47.9k \\
\texttt{seed-diverse} & 0.233 & 0.628 & 0.431 & -- & 26.8k / 27.0k & 47.0k / 47.4k \\
\texttt{seed-minibatch} & 0.194 & 0.611 & 0.403 & -- & 26.6k / 26.5k & 47.1k / 47.6k \\
\texttt{seed-clustergold} & 0.250 & 0.681 & 0.466 & -- & 41.4k / 41.8k & 61.8k / 61.5k \\
\texttt{parent-first} & 0.372 & 0.653 & 0.513 & \texttt{{[}6,11,4,11,7,9,4,4,0,0{]}} (5.6 / 12) & 38.7k / 37.3k & 52.5k / 55.3k \\
\texttt{parent-diverse} & 0.403 & 0.656 & 0.530 & -- & 37.8k / 37.3k & 53.0k / 55.0k \\
\texttt{parent-minibatch} & 0.361 & 0.642 & 0.502 & -- & 35.4k / 35.6k & 53.9k / 54.7k \\
\texttt{parent-clustergold} & 0.400 & 0.667 & 0.534 & -- & 55.4k / 56.4k & 69.8k / 71.4k \\
\midrule
\multicolumn{7}{@{}l@{}}{\textbf{Panel B: GEPA Baseline (Batch A.4) and Discovery-Motivated \texttt{carry-facts} Follow-Up (Batch A.5, \(n=10\))}} \\
\midrule
\texttt{GEPA\ mb3} (No early stop) & 0.664 & 0.633 & 0.649 & Median 9.0, Mean 9.0 / 12 & 106.9k / 107.3k & 121.6k / 121.9k \\
\texttt{GEPA\ mb3\ +\ stop} & 0.661 & 0.644 & 0.653 & -- & 102.3k / 105.3k & 114.9k / 118.5k \\
\texttt{seed-first\ +\ carry} & 0.317 & 0.614 & 0.466 & \texttt{{[}4,4,4,4,4,4,4,4,4,4{]}} (4.0 / 12) & 34.6k / 34.3k & 49.0k / 49.1k \\
\texttt{parent-first\ +\ carry} & \textbf{0.856} & 0.592 & \textbf{0.724} & \texttt{{[}11,12,12,11,12,11,12,12,12,12{]}} (11.7 / 12) & \textbf{45.8k} / 51.0k & 55.5k / 55.7k \\
\bottomrule
\end{tabular}%
}
\end{center}

\textbf{Table 4:} Held-out accuracy, intent-label retention in the
search-best prompt, and search tokens per run (median / mean across 10
runs) on \texttt{banking77} and \texttt{bbh\_ste}.

In the pre-registered grid (Table 4, Panel A), parent inheritance
increases pooled public-benchmark accuracy by
\(\Delta_{\text{lineage}} = +0.088\), driven by \texttt{banking77} at
+0.158 while \texttt{bbh\_ste} changes by only +0.017. However,
\texttt{GEPA\ mb3} initially outperformed our baseline
\texttt{parent-first} loop on \texttt{banking77}, scoring 0.664 against
0.372, while matching it on \texttt{bbh\_ste} at 0.633 versus 0.653.

Inspecting the Batch A.4 \texttt{banking77} logs identified two reasons
for this gap. First, because the seed instruction
\texttt{"Classify\ the\ banking\ customer\ query."} contains none of the
12 \texttt{snake\_case} intent labels, first-round edits that only
adjusted formatting scored 0.000 and were rejected under
strict-improvement tie-breaking; only 3 of 10 \texttt{parent-first} runs
improved in round 1, and 2 runs stayed at 0.000 across all 8
evaluations. Second, each run's held-out accuracy closely tracked how
many of the 12 ground-truth intent names appeared in its final prompt,
averaging 5.6 out of 12 for \texttt{parent-first} versus 9.0 out of 12
for \texttt{GEPA\ mb3} (Table 4). Comparing the meta-prompts showed why:
GEPA 0.0.7's reflection prompt explicitly instructs the model to
\emph{``identify all niche and domain specific factual information about
the task and include it in the instruction,''} whereas our proposer
objective had no such instruction.

\begin{figure}
\centering
\includegraphics{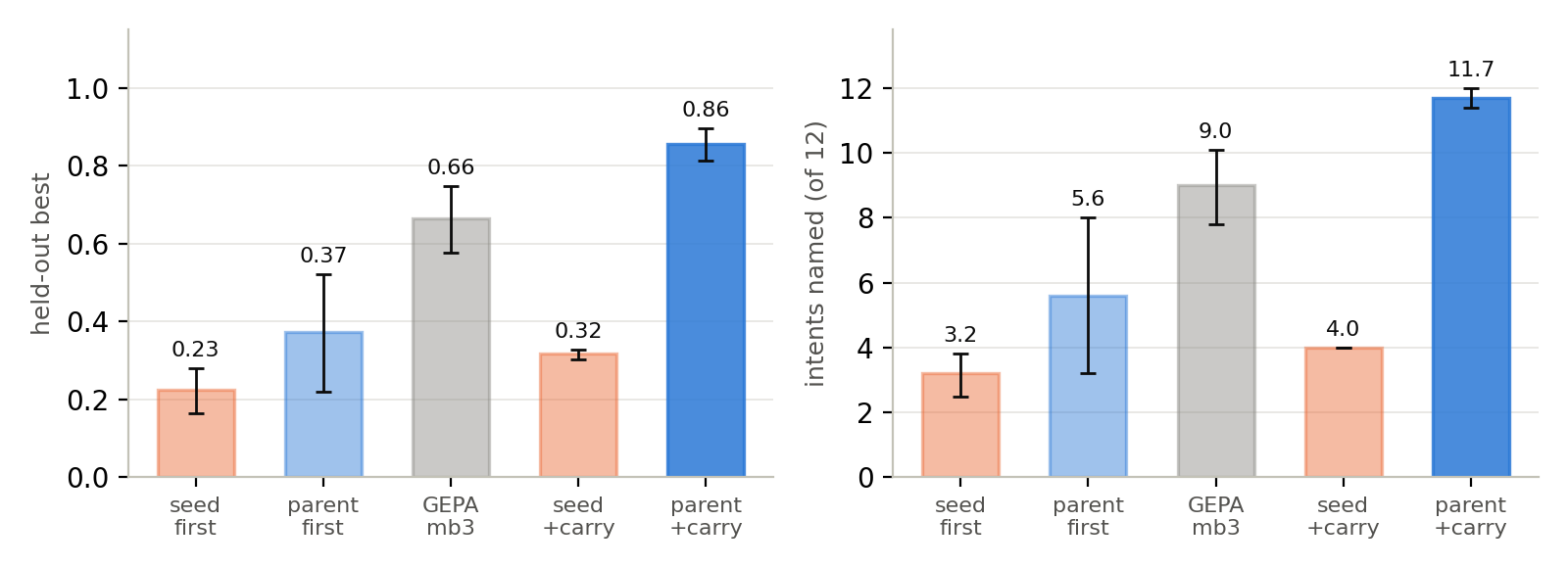}
\caption{Held-out accuracy (left) and number of ground-truth Banking77
intents named in the search-best prompt out of 12 (right) across
optimization conditions (\texttt{gpt-4.1-mini}, \(n=10\) runs per bar,
error bars show 95\% bootstrap confidence intervals). Orange bars reset
to the seed each round (\texttt{seed}); blue bars build on the best
parent (\texttt{parent}); grey bars show \texttt{GEPA\ mb3} at matched
task-call caps; \texttt{+carry} appends the single-sentence
fact-carrying instruction.}
\end{figure}

To test whether explicit fact copying accounts for this gap, we
registered a follow-up batch (Appendix A.5) that appended one sentence,
\texttt{CARRY\_FACTS} (Appendix B.1), to the proposer objective:
\emph{``The task model sees only this prompt, never the feedback: copy
every task-specific fact the feedback reveals (for example the exact
expected labels or formats) into the prompt.''} On \texttt{banking77}
(Table 4, Panel B and Figure 2), appending \texttt{CARRY\_FACTS} raised
\texttt{parent-first} accuracy from 0.372 to 0.856, with all 10 runs
improving in round 1 and surpassing \texttt{GEPA\ mb3} by +0.192 at 43\%
of its median token cost, using 45.8k versus 106.9k tokens. This gain
requires parent inheritance: \texttt{seed-first\ +\ carry} observes the
same 4 seed failures every round, writes 4 intents into the prompt in
every run, and scores 0.317, whereas \texttt{parent-first\ +\ carry}
accumulates 11 to 12 intents across rounds with a mean of 11.7 out of
12. Conversely, on \texttt{bbh\_ste}, where all six option labels are
already given in the seed, appending \texttt{CARRY\_FACTS} reduced
\texttt{parent-first} accuracy from 0.653 to 0.592, a change of -0.061
that fails non-inferiority test \(\mathcal{H}_{F3}\). Copying feedback
facts into the prompt helps when label inventories must be discovered
from feedback, and slightly hurts when the specification is already
complete in the seed.

\hypertarget{interaction-between-lineage-and-feedback-diversity-mathcalh_c2}{%
\subsection{\texorpdfstring{Interaction Between Lineage and Feedback
Diversity
(\(\mathcal{H}_{C2}\))}{Interaction Between Lineage and Feedback Diversity (\textbackslash mathcal\{H\}\_\{C2\})}}\label{interaction-between-lineage-and-feedback-diversity-mathcalh_c2}}

When resetting to the seed, replacing the first four failures with four
farthest-point diverse failures raises 7-task accuracy by +0.061 on
\texttt{gpt-4.1-mini} (Table 2B), led by \texttt{labels},
\texttt{duration}, and \texttt{replies}. Under parent inheritance, the
difference between \texttt{diverse} and \texttt{first} is -0.007,
yielding a positive attenuation interaction of
\(I_{\text{diverse}} = +0.068\) that supports \(\mathcal{H}_{C2}\) and
rejects \(\mathcal{H}_{R1}\). This interaction replicates on
\texttt{claude-haiku-5-5} across the five non-ceiling tasks at
\(I_{\text{diverse}} = +0.122\) (Table 3, Panel A). At the task level,
\texttt{size} is the only task with a materially negative
\(I_{\text{diverse}}\) on \texttt{gpt-4.1-mini} at -0.088, while
\texttt{routing} is the only non-ceiling task where
\texttt{seed-diverse} scores below \texttt{seed-first} on both models
because surface edit distance selects two \texttt{fulfillment} tickets
in round 1, covering 3 queues instead of 4.

\hypertarget{unshown-category-coverage-mathcalh_c1-and-cross-round-exposure}{%
\subsection{\texorpdfstring{Unshown-Category Coverage
(\(\mathcal{H}_{C1}\)) and Cross-Round
Exposure}{Unshown-Category Coverage (\textbackslash mathcal\{H\}\_\{C1\}) and Cross-Round Exposure}}\label{unshown-category-coverage-mathcalh_c1-and-cross-round-exposure}}

Panel A of Table 5 tests whether showing a narrow set of failures under
seed resetting crowds out pre-trained capabilities on unshown
categories.

\begin{center}
\small\setlength{\tabcolsep}{4.5pt}\renewcommand{\arraystretch}{1.18}
\resizebox{\ifdim\width>\textwidth\textwidth\else\width\fi}{!}{%
\begin{tabular}{@{}l>{\raggedright\arraybackslash}p{2.3cm}>{\raggedright\arraybackslash}p{3.5cm}cccc>{\raggedright\arraybackslash}p{3.4cm}@{}}
\toprule
\textbf{Task Family} & \textbf{Shown Categories in \texttt{seed-first}} & \textbf{Unshown Failing Categories (\(\mathcal{U}_\tau\))} & \parbox[b]{2.25cm}{\centering\textbf{\texttt{seed-none} Coverage}} & \parbox[b]{2.25cm}{\centering\textbf{\texttt{seed-first} Coverage}} & \parbox[b]{2.25cm}{\centering\textbf{\texttt{seed-diverse} Coverage}} & \parbox[b]{2.25cm}{\centering\textbf{\texttt{seed-cluster} Coverage}} & \textbf{Contrast (\texttt{seed-first} \(-\) \texttt{seed-none}, 95\% CI)} \\
\midrule
\texttt{labels} & \texttt{iso\_date}, \texttt{us\_date} & \texttt{email}, \texttt{eu\_date}, \texttt{hex\_color}, \texttt{ipv4}, \texttt{ipv6}, \texttt{mac\_address}, \texttt{semver} & 0.133 & 0.151 & 0.414 & 0.545 & +0.018 {[}-0.031, +0.067{]} \\
\texttt{duration} & \texttt{days/weeks}, \texttt{number}, \texttt{units} & \texttt{clock}, \texttt{iso}, \texttt{words} & 0.058 & 0.003 & 0.009 & -- & \textbf{-0.055} {[}-0.082, -0.033{]} \\
\texttt{duration} (\texttt{seed-iso} / \texttt{seed-random}) & \texttt{iso} only / 4 random failures & \texttt{clock}, \texttt{iso}, \texttt{words} & 0.058 & -- & -- & -- & \texttt{iso}: 0.009 (\textbf{-0.048} {[}-0.079, -0.021{]}) / \texttt{rand}: 0.118 (+0.061 {[}-0.003, +0.121{]}) \\
\texttt{size} & \texttt{bytes}, \texttt{si} & \texttt{iec}, \texttt{invalid}, \texttt{letter}, \texttt{separator}, \texttt{words} & 0.431 & 0.318 & 0.345 & 0.945 & \textbf{-0.113} {[}-0.144, -0.082{]} \\
\bottomrule
\end{tabular}%
}
\end{center}

\textbf{Table 5A:} Proposal-level coverage of unshown failing categories
in the \texttt{seed} row (\(\mathcal{H}_{C1}\), \texttt{gpt-4.1-mini},
\(n=10\) runs per cell).

\begin{center}
\small\setlength{\tabcolsep}{4.5pt}\renewcommand{\arraystretch}{1.18}
\resizebox{\ifdim\width>\textwidth\textwidth\else\width\fi}{!}{%
\begin{tabular}{@{}lccccccc@{}}
\toprule
\textbf{Task} & \parbox[b]{2.25cm}{\centering\textbf{Union of Search Failures (\texttt{seed-first})}} & \parbox[b]{2.25cm}{\centering\textbf{Union of Search Failures (\texttt{parent-first})}} & \parbox[b]{2.25cm}{\centering\textbf{Parent-to-Child Retention (\texttt{parent-first})}} & \parbox[b]{2.25cm}{\centering\textbf{Mean Best Chars (\texttt{seed} / \texttt{parent})}} & \parbox[b]{2.25cm}{\centering\textbf{Mean Search Tokens (\texttt{seed} / \texttt{parent})}} & \parbox[b]{2.25cm}{\centering\textbf{Solved Runs (\texttt{seed} / \texttt{parent})}} & \parbox[b]{2.25cm}{\centering\textbf{Held-Out Best (\texttt{seed-first} / \texttt{parent-first})}} \\
\midrule
\texttt{labels} & 0.17 (4 / 24) & \textbf{0.54} & 0.70 & 335 / 579 & 27.3k / 25.0k & 0/10 vs.~7/10 & 0.408 / 0.917 \\
\texttt{dates} & 0.17 (4 / 24) & 0.17 & 1.00 & 188 / 191 & 4.5k / 4.5k & 10/10 vs.~10/10 & 1.000 / 1.000 \\
\texttt{routing} & 0.25 (4 / 16) & \textbf{0.61} & 0.91 & 327 / 1512 & 17.7k / 27.7k & 0/10 vs.~9/10 & 0.500 / 0.925 \\
\texttt{replies} & 0.25 (4 / 16) & \textbf{0.86} & 0.83 & 792 / 2314 & 39.7k / 35.2k & 0/10 vs.~10/10 & 0.294 / 0.981 \\
\texttt{scone} & 0.11 (4 / 36) & 0.12 & 0.97 & 280 / 481 & 46.1k / 54.1k & 0/10 vs.~2/10 & 0.899 / 0.910 \\
\texttt{duration} & 0.17 (4 / 24) & \textbf{0.51} & 0.95 & 1139 / 4296 & 10.3k / 15.9k & 0/10 vs.~6/10 & 0.562 / 0.963 \\
\texttt{size} & 0.17 (4 / 24) & \textbf{0.46} & 0.83 & 1782 / 2504 & 13.0k / 8.2k & 0/10 vs.~8/10 & 0.767 / 0.929 \\
\bottomrule
\end{tabular}%
}
\end{center}

\textbf{Table 5B:} Cross-round search-set failure exposure,
parent-to-child retention, mean candidate length, and mean search tokens
per run in \texttt{first-k} cells (\texttt{gpt-4.1-mini}, \(n=10\) runs
per cell).

On both code synthesis tasks, rewriting the 4-line seed function with
zero feedback in \texttt{seed-none} often produces general parsers that
handle several unshown formats, reaching 0.058 unshown coverage on
\texttt{duration} and 0.431 on \texttt{size}. When \texttt{seed-first}
shows four unit-suffix or SI failures, the proposer specializes its code
to those displayed formats and drops handlers for unshown categories,
lowering unshown coverage by -0.055 on \texttt{duration} and -0.113 on
\texttt{size}, and lowering held-out accuracy on \texttt{duration} below
\texttt{seed-none} on both \texttt{gpt-4.1-mini} and
\texttt{claude-haiku-5-5} (Tables 2A and 3). The causal intervention on
\texttt{duration} confirms this mechanism: showing only ISO-8601
failures in \texttt{seed-iso} reduces unshown coverage by -0.048,
whereas showing 4 random failures per round in \texttt{seed-random}
changes it by +0.061. On \texttt{labels}, where the zero-feedback prior
rarely guesses hidden \texttt{snake\_case} names, no drop occurs, so
\(\mathcal{H}_{C1}\) holds on 2 of 3 tested tasks.

Panel B of Table 5 shows how parent inheritance avoids this trap. While
\texttt{seed-first} shows the exact same 4 failures every round,
covering only 11\% to 25\% of \(\mathcal{D}_{\text{search}}\),
\texttt{parent-first} retains 70\% to 95\% of parent-solved examples
when generating children and filters solved examples from the failure
list, exposing 46\% to 86\% of \(\mathcal{D}_{\text{search}}\) across
rounds on the five non-ceiling tasks. Because early stopping halts
search once \(\mathcal{D}_{\text{search}}\) is solved,
\texttt{parent-first} also uses fewer or equal mean search tokens on
four of the seven tasks despite producing longer candidates.

\hypertarget{full-set-error-clustering-with-and-without-lineage}{%
\subsection{Full-Set Error Clustering With and Without
Lineage}\label{full-set-error-clustering-with-and-without-lineage}}

In Batch A.1, the initial \texttt{cluster} arm underperformed
\texttt{first} in the parent row by -0.082 and lost to the best
alternative disclosure scheme in both rows, refuting
\(\mathcal{H}_{R2}\) (Table 7). Leave-one-task-out analysis (Table 8)
traced this deficit to \texttt{replies}, where \texttt{parent-cluster}
scored 0.519 against 0.981 for \texttt{parent-first} because
\texttt{CLUSTER\_PROMPT} asked for a representative wrong answer without
requiring the expected answer and dropped the grading rubric in 125 of
156 summaries. In Batch A.2, \texttt{clustergold} required each cluster
line to include the input, wrong answer, and full expected answer,
restoring 100\% rubric retention on \texttt{replies} across all 65
summaries and raising \texttt{parent-clustergold} on \texttt{replies} to
0.869 (Table 2A).

\begin{center}
\small\setlength{\tabcolsep}{4.5pt}\renewcommand{\arraystretch}{1.18}
\resizebox{\ifdim\width>\textwidth\textwidth\else\width\fi}{!}{%
\begin{tabular}{@{}lccccccccc@{}}
\toprule
\parbox[b]{2.55cm}{\raggedright\textbf{Proposal Round (\(t\))}} & \textbf{\texttt{seed-first}} & \textbf{\texttt{seed-diverse}} & \textbf{\texttt{seed-cgold}} & \parbox[b]{1.95cm}{\centering\textbf{\texttt{seed} (\texttt{cgold} \(-\) \texttt{first})}} & \textbf{\texttt{parent-first}} & \parbox[b]{1.95cm}{\centering\textbf{\texttt{parent-diverse}}} & \textbf{\texttt{parent-cgold}} & \parbox[b]{1.95cm}{\centering\textbf{\texttt{parent} (\texttt{cgold} \(-\) \texttt{first})}} & \parbox[b]{1.95cm}{\centering\textbf{3 Prompt / 2 Code (\texttt{parent} Diff.)}} \\
\midrule
\(t = 1\) (Edits \(c_0\)) & 0.446 & 0.550 & 0.659 & \textbf{+0.213} {[}+0.149, +0.272{]} & 0.487 & 0.533 & 0.633 & \textbf{+0.146} {[}+0.083, +0.205{]} & +0.183 / +0.090 \\
\(t = 2\) & 0.518 & 0.582 & 0.707 & \textbf{+0.189} {[}+0.136, +0.238{]} & 0.669 & 0.779 & 0.842 & \textbf{+0.173} {[}+0.118, +0.219{]} & +0.209 / +0.119 \\
\(t = 4\) & 0.564 & 0.628 & 0.747 & \textbf{+0.183} {[}+0.137, +0.222{]} & 0.946 & 0.926 & 0.929 & -0.017 {[}-0.067, +0.023{]} & -0.040 / +0.019 \\
\(t = 7\) & 0.587 & 0.650 & 0.767 & \textbf{+0.180} {[}+0.132, +0.219{]} & 0.971 & 0.983 & 0.956 & -0.015 {[}-0.065, +0.024{]} & -0.038 / +0.019 \\
\bottomrule
\end{tabular}%
}
\end{center}

\textbf{Table 5C:} Anytime search-set accuracy across the five
non-ceiling tasks on \texttt{gpt-4.1-mini} at proposal steps
\(t \in \{1, 2, 4, 7\}\), with 95\% bootstrap intervals for
\texttt{clustergold} \(-\) \texttt{first} (\texttt{cgold} \(-\)
\texttt{first}) and separate \texttt{parent} differences for the 3
prompt tasks with \(B=8\) and 2 code tasks with \(B=12\).

With expected answers preserved, the benefit of full-set clustering
depends on both lineage and the optimization round (Tables 2B and 5C):

\begin{itemize}
\tightlist
\item
  \textbf{Under seed resetting:} \texttt{seed-clustergold} is the
  strongest seed-row method on both \texttt{gpt-4.1-mini}, where it
  reaches 0.789 and beats \texttt{seed-first} by +0.156, and
  \texttt{claude-haiku-5-5}, where it reaches 0.823 and beats
  \texttt{seed-first} by +0.179.
\item
  \textbf{Under parent inheritance:} At terminal evaluation,
  \texttt{parent-clustergold} and \texttt{parent-first} show no
  detectable difference on \texttt{gpt-4.1-mini} at -0.018 or on
  \texttt{claude-haiku-5-5} at -0.008, yielding positive attenuation
  interactions of \(I_{\text{clustergold}} = +0.174\) and \(+0.187\), as
  well as \(+0.916\) log-odds units on an exploratory logit scale, while
  \texttt{clustergold} scores lower by -0.057 under the \texttt{gpt-4.1}
  proposer. Table 5C shows how this convergence unfolds across rounds:
  at \(t=1\), where both rows edit \(c_0\), and at \(t=2\), the first
  parent-inheritance step, \texttt{clustergold} leads \texttt{first} by
  +0.146 and +0.173 in the parent row. By \(t=4\), sequential error
  exposure under \texttt{parent-first} closes the gap on both prompt and
  code tasks, reaching 0.946 versus 0.929, whereas \texttt{seed-first}
  remains 0.180 behind \texttt{seed-clustergold} at \(t=7\). Moreover,
  the extra clusterer call increases median search tokens on five of
  seven tasks, reaching up to 2.08\(\times\) on \texttt{routing} in
  \texttt{parent} at 58.5k versus 28.2k tokens and 2.17\(\times\) in
  \texttt{seed} at 38.1k versus 17.6k tokens.
\item
  \textbf{Clustered failures vs.~random mixed minibatches:} Under parent
  inheritance, \texttt{parent-clustergold} still beats
  \texttt{parent-minibatch} by +0.043 on \texttt{gpt-4.1-mini} and
  +0.086 on Haiku because \texttt{parent-minibatch} wastes display slots
  on already-passing examples and scores below \texttt{parent-first} by
  -0.061 and -0.093.
\end{itemize}

\hypertarget{system-level-search-efficiency-and-token-decomposition-against-gepa-mathcalh_c3}{%
\subsection{\texorpdfstring{System-Level Search Efficiency and Token
Decomposition Against GEPA
(\(\mathcal{H}_{C3}\))}{System-Level Search Efficiency and Token Decomposition Against GEPA (\textbackslash mathcal\{H\}\_\{C3\})}}\label{system-level-search-efficiency-and-token-decomposition-against-gepa-mathcalh_c3}}

Table 6 compares \texttt{evolve\ (parent-first)} against
\texttt{GEPA\ 0.0.7} with and without early stopping at matched maximum
task-model call caps across all seven prompt tasks.

\begin{center}
\small\setlength{\tabcolsep}{4.5pt}\renewcommand{\arraystretch}{1.18}
\resizebox{\ifdim\width>\textwidth\textwidth\else\width\fi}{!}{%
\begin{tabular}{@{}lcccc>{\centering\arraybackslash}p{3.9cm}c@{}}
\toprule
\textbf{Task Family} & \parbox[b]{2.25cm}{\centering\textbf{Solved Runs (\texttt{evolve} / \texttt{GEPA+stop} / \texttt{GEPA})}} & \parbox[b]{2.25cm}{\centering\textbf{Held-Out Best (\texttt{evolve} / \texttt{GEPA+stop} / \texttt{GEPA})}} & \parbox[b]{2.25cm}{\centering\textbf{Conditional Median Tokens to Solve (\texttt{evolve} / \texttt{GEPA+stop} / \texttt{GEPA})}} & \parbox[b]{2.25cm}{\centering\textbf{Tokens-to-Solve Ratio (\texttt{GEPA+stop} / \texttt{evolve}, 95\% CI)}} & \textbf{Unconditional Search Tokens / Run, Median / Mean (\texttt{evolve} \(\vert\) \texttt{GEPA+stop} \(\vert\) \texttt{GEPA})} & \parbox[b]{2.25cm}{\centering\textbf{GEPA (\texttt{mb3}) Token Share: Full Re-Eval / Minibatch / Reflection}} \\
\midrule
\texttt{labels} & 7/10 / 6/10 / 6/10 & 0.917 / 0.908 / 0.917 & 21.4k / 71.6k / 67.7k & \textbf{3.34\(\times\)} {[}2.23, 4.49{]} & 23.1k / 25.0k \(\vert\) 78.1k / 78.8k \(\vert\) 86.8k / 88.9k & 67\% / 27\% / 7\% \\
\texttt{dates} & 10/10 / 10/10 / 9/10 & 1.000 / 0.988 / 0.992 & 4.4k / 11.5k / 11.9k & \textbf{2.59\(\times\)} {[}2.30, 4.84{]} & 4.4k / 4.5k \(\vert\) 11.5k / 22.2k \(\vert\) 57.7k / 62.3k & 21\% / 77\% / 2\% \\
\texttt{routing} & 9/10 / 2/10 / 1/10 & 0.925 / 0.842 / 0.750 & 27.3k / 53.7k / 68.0k & Ineligible (\(<3\) GEPA solved) & 28.2k / 27.7k \(\vert\) 72.9k / 65.9k \(\vert\) 68.1k / 68.6k & 45\% / 39\% / 15\% \\
\texttt{replies} & 10/10 / 2/10 / 2/10 & 0.981 / 0.850 / 0.850 & 34.3k / 79.6k / 81.1k & Ineligible (\(<3\) GEPA solved) & 34.3k / 35.2k \(\vert\) 88.3k / 94.6k \(\vert\) 97.1k / 94.2k & 61\% / 28\% / 11\% \\
\texttt{scone} & 2/10 / 4/10 / 4/10 & 0.910 / 0.899 / 0.918 & 47.5k / 50.6k / 97.8k & Ineligible (\(<3\) \texttt{evolve} solved) & 53.5k / 54.1k \(\vert\) 155.2k / 130.3k \(\vert\) 149.2k / 152.4k & 56\% / 38\% / 5\% \\
\texttt{banking77} (Base / \texttt{+carry}) & 0/10 (0/10) / 0/10 / 0/10 & 0.372 (\textbf{0.856}) / 0.661 / 0.664 & -- / -- / -- & Ineligible (0 solved runs) & 38.7k / 37.3k (45.8k / 51.0k) \(\vert\) 102.3k / 105.3k \(\vert\) 106.9k / 107.3k & 74\% / 18\% / 8\% \\
\texttt{bbh\_ste} (Base / \texttt{+carry}) & 0/10 (0/10) / 0/10 / 0/10 & 0.653 (0.592) / 0.644 / 0.633 & -- / -- / -- & Ineligible (0 solved runs) & 52.5k / 55.3k (55.5k / 55.7k) \(\vert\) 114.9k / 118.5k \(\vert\) 121.6k / 121.9k & 58\% / 30\% / 12\% \\
\bottomrule
\end{tabular}%
}
\end{center}

\textbf{Table 6:} System-level comparison of
\texttt{evolve\ (parent-first)} against \texttt{GEPA\ 0.0.7}
(\texttt{minibatch=3}, with and without early stopping) at matched
maximum task-model call caps (\(n=10\) runs per cell,
\texttt{gpt-4.1-mini}).

\texttt{evolve\ (parent-first)} solves the search set at least as often
as \texttt{GEPA\ +\ stop} on four of the five primary prompt tasks,
while \texttt{GEPA\ +\ stop} solves more often on \texttt{scone}. Among
solved runs on the two tasks meeting the \(\ge 3/10\) eligibility rule,
\texttt{labels} and \texttt{dates}, \texttt{GEPA\ +\ stop} median
tokens-to-solve is 3.34\(\times\) and 2.59\(\times\) that of
\texttt{parent-first} (Table 6). On \texttt{dates}, two right-tail
\texttt{GEPA\ +\ stop} runs at 30.4k and 100.5k tokens widen the upper
bootstrap bound to 4.84\(\times\), while the 5-task refutation test
\(\mathcal{H}_{R4}\) is not evaluable as registered because three tasks
had fewer than 3 solved runs in one arm. Across all seven prompt tasks,
unconditional median search tokens per run for \texttt{parent-first} are
2.2\(\times\) to 3.4\(\times\) lower than for \texttt{GEPA\ +\ stop}.
Call-type accounting in the last column of Table 6 shows why: proposer
reflection accounts for only 2\% to 15\% of GEPA's tokens, whereas full
search-set validations and minibatch checks consume 85\% to 98\%.

\hypertarget{boundary-conditions-robustness-and-negative-findings}{%
\subsection{Boundary Conditions, Robustness, and Negative
Findings}\label{boundary-conditions-robustness-and-negative-findings}}

\begin{enumerate}
\def\labelenumi{\arabic{enumi}.}
\tightlist
\item
  \textbf{Unresolved disclosure contrasts on the public benchmarks
  (\(\mathcal{H}_{P2}\text{--}\mathcal{H}_{P5}\)):} While parent
  inheritance improves pooled accuracy on \texttt{banking77} and
  \texttt{bbh\_ste}, the disclosure interactions and parent-row
  contrasts on these two tasks have wide confidence intervals centered
  near zero due to run-to-run variance on baseline \texttt{banking77}
  and flat disclosure performance on \texttt{bbh\_ste} (Table 7).
\item
  \textbf{Task-specificity of \texttt{CARRY\_FACTS}
  (\(\mathcal{H}_{F3}\)):} Appending \texttt{CARRY\_FACTS} lowers
  \texttt{parent-first} accuracy on \texttt{bbh\_ste} by -0.061, where
  all option labels are already present in the seed.
\item
  \textbf{Task-level exceptions:} On \texttt{routing}, surface
  edit-distance \texttt{diverse} sampling scores below \texttt{first}
  under seed resetting on both models; under parent inheritance on
  \texttt{gpt-4.1-mini}, \texttt{parent-clustergold} scores below
  \texttt{parent-first} on \texttt{replies} by -0.112 and on
  \texttt{routing} by -0.083. On \texttt{scone}, all cells sit near
  ceiling between 0.889 and 0.918.
\item
  \textbf{Leave-one-task-out stability:} Dropping any single task (Table
  8), \(\Delta_{\text{lineage}}\) stays between +0.196 and +0.278 on
  \texttt{gpt-4.1-mini} and between +0.168 and +0.257 on Haiku,
  \(I_{\text{diverse}}\) stays between +0.046 and +0.094, and
  \texttt{parent-minibatch} \(-\) \texttt{parent-first} stays between
  -0.042 and -0.069.
\end{enumerate}

\hypertarget{discussion}{%
\section{Discussion}\label{discussion}}

\textbf{Why parent inheritance substitutes for wide single-round
disclosure.} Under seed resetting, the proposer has no persistent state,
so a narrow feedback sample can anchor the rewrite on displayed formats
and crowd out zero-feedback priors on unshown categories (Table 5, Panel
A). Even when a full-set summary in \texttt{seed-clustergold} exposes
all failing categories, synthesizing every rule from scratch in a single
edit reaches only 0.723 across the five feedback-dependent tasks. Under
parent inheritance, the candidate text retains rules discovered in
earlier rounds, preserving 70\% to 95\% of solved cases, while full
search-set evaluation removes those solved cases from future failure
lists, shifting \texttt{first-k} disclosure across 46\% to 86\% of
\(\mathcal{D}_{\text{search}}\) over a run (Table 5, Panel B).
Consequently, while \texttt{clustergold} speeds up learning in rounds 1
and 2, \texttt{parent-first} catches up by round 4 (Table 5C). On tasks
such as \texttt{banking77} where target labels are hidden in feedback,
this accumulation works only if the proposer writes revealed domain
facts into the candidate prompt via \texttt{CARRY\_FACTS}
\citep{shi2025integrating}; when all options are already in the seed, as
on \texttt{bbh\_ste}, copying training facts slightly hurts
generalization. This contrast parallels selective LLM-guided
regularization in recommendation \citep{shi2026selective}, where
applying LLM guidance indiscriminately degrades performance on
well-specified instances while selective intervention targeted to
under-specified cases yields consistent gains.

\textbf{Implications for evaluating reflective optimizers.} Because
recent reflection-scope methods such as ESPO \citep{liu2026espo}, RPT
\citep{bayat2026reflective}, and AdaGEPA \citep{chen2026adagepa} are
tested only inside lineage-preserving loops, their reported gains
measure residual improvements on top of parent inheritance. In our grid,
\texttt{parent-clustergold} beats \texttt{parent-minibatch} by +0.043 on
\texttt{gpt-4.1-mini} and +0.086 on Haiku, but does not beat
\texttt{parent-first} at termination. Gains over default GEPA in
short-budget settings can therefore come from replacing mixed pass/fail
minibatches with pure failure feedback rather than from clustering
itself. Consistent with recent evaluation audits of self-evolving
systems \citep{yang2026selfevolvingrecsys}, future evaluations of
reflection-scope or proposer-upgrade methods should include three
controls: a lineage-off \texttt{seed} baseline row alongside the
\texttt{parent} row, a zero-feedback \texttt{seed-none} baseline to
detect crowding-out, and call-type token accounting separating
reflection from validation.

\textbf{Practical guidelines.} When tuning prompts or code under short
evaluation budgets, practitioners should build on the best parent
candidate while monitoring for zero-score stalls on tasks with hidden
labels, show raw failing examples with full expected answers instead of
random pass/fail minibatches, instruct the proposer to copy revealed
task rules or labels into the prompt when they are absent from the seed,
and reserve full-set error clustering for one-shot repair, parallel
best-of-\(N\) sampling, or the first two rounds of search.

\hypertarget{limitations}{%
\section{Limitations}\label{limitations}}

Several constraints bound our conclusions. First, five of the seven
primary tasks and \texttt{replies} are constructed by withholding rules
from the seed, which favors cross-round accumulation; on the two public
extension tasks, \texttt{banking77} and \texttt{bbh\_ste}, the pooled
lineage effect is smaller at +0.088 versus +0.225 across the same four
disclosure schemes on the primary tasks, and the disclosure interactions
\(\mathcal{H}_{P2}\text{--}\mathcal{H}_{P5}\) are not statistically
resolved. Grouping search examples by category on constructed tasks also
reduces round-1 category diversity under \texttt{first}, and on larger
tasks where \(B\) rounds of \(k=4\) failures cannot cover the error
space, wider per-round disclosure may retain an advantage beyond round
4. Second, while \texttt{parent-clustergold} versus
\texttt{seed-clustergold} shows that parent inheritance adds +0.139
overall and +0.198 on non-ceiling tasks even when all failure categories
are summarized every round, \texttt{seed-clustergold} compresses
failures into 3 to 7 patterns; fully separating text inheritance from
residual-error exposure under raw feedback would require controls that
reset to \(c_0\) while showing all raw failures or the cumulative union
of parent-trajectory failures. Third, our comparison with GEPA 0.0.7 is
system-level across parent selection, minibatch gating, minibatch size,
and reflection wording. Fourth, \texttt{carry-facts} was designed after
inspecting Batch A.4 \texttt{banking77} logs and tested on two public
benchmarks, helping \texttt{banking77} by +0.483 and hurting
\texttt{bbh\_ste} by -0.061. Fifth, \texttt{banking77} was subsampled to
12 intents and \texttt{bbh\_ste} to 198 examples, with held-out sets of
12 to 72 examples across tasks, to fit the 8-evaluation budget and
300-character feedback window. Sixth, \texttt{dates} and \texttt{scone}
start near ceiling. Seventh, \texttt{cluster} and \texttt{clustergold}
test only the Diagnose stage of ESPO \citep{liu2026espo}, and
\texttt{diverse} uses surface Levenshtein distance. Eighth, experiments
use two model families, \texttt{gpt-4.1-mini} / \texttt{gpt-4.1} and
\texttt{claude-haiku-5-5} with default adaptive thinking, and a single
\texttt{gpt-4.1-mini} judge on \texttt{replies}. Finally, batches ran
sequentially over two days, and three intermediate batches interrupted
by API quota exhaustion totaling 267 aborted run attempts were
quarantined and re-run from scratch under pre-registered rules (Appendix
A).

\hypertarget{appendix-a-pre-registered-protocols-hypothesis-inventory-and-robustness-tables}{%
\section{Appendix A: Pre-Registered Protocols, Hypothesis Inventory, and
Robustness
Tables}\label{appendix-a-pre-registered-protocols-hypothesis-inventory-and-robustness-tables}}

All protocols, sample sizes, estimands, and decision rules were
registered in \texttt{preregistration-zh.md} and
\texttt{experiments.json} before executing each batch and translated in
\texttt{preregistration-en.md}, with frozen code commits and SHA-256
data hashes archived in the repository:

\begin{itemize}
\tightlist
\item
  \textbf{A.1 Main Factorial Grid (\texttt{gpt-4.1-mini}, 750 runs,
  commit \texttt{d5ab8923}):} Registered
  \(P \in \{\text{seed}, \text{parent}\} \times F \in \{\text{none}, \text{first}, \text{diverse}, \text{minibatch}, \text{cluster}\}\)
  across 7 tasks with \(n=10\), \texttt{duration} interventions
  \texttt{seed-iso} and \texttt{seed-random}, and \texttt{GEPA\ mb3} /
  \texttt{GEPA\ mb3\ +\ stop}, defining \(\Delta_{\text{lineage}}\),
  \(\mathcal{H}_{C1}\text{--}\mathcal{H}_{C3}\), and
  \(\mathcal{H}_{R1}\text{--}\mathcal{H}_{R4}\).
\item
  \textbf{A.2 Target-Preserving Cluster Sensitivity
  (\texttt{sens-clustergold}, \texttt{gpt-4.1-mini}, 140 runs):}
  Registered after auditing rubric omission in Batch A.1
  \texttt{replies-cluster}, testing \(\mathcal{H}_{S1\text{-gold}}\) and
  \(\mathcal{H}_{S2\text{-gold}}\) with \texttt{CLUSTER\_GOLD\_PROMPT}.
\item
  \textbf{A.3 Second-Model Replication (\texttt{rep-haiku},
  \texttt{claude-haiku-5-5}, 350 runs):} Registered 5 non-ceiling tasks
  across 7 cells, namely \texttt{seed-none} and
  \(\{\text{seed}, \text{parent}\} \times \{\text{first}, \text{minibatch}, \text{clustergold}\}\),
  testing \(\mathcal{H}_{H1}\text{--}\mathcal{H}_{H3}\).
\item
  \textbf{A.4 Public Benchmarks, Strong Proposer, and Haiku Diverse (470
  runs):} Registered 2 public tasks, \texttt{banking77} and
  \texttt{bbh\_ste}, across 9 factorial cells plus 2 GEPA arms
  (\(\mathcal{H}_{P1}\text{--}\mathcal{H}_{P5}\)), 3 tasks with
  \texttt{gpt-4.1} proposer
  (\(\mathcal{H}_{S1\text{-prop}}\text{--}\mathcal{H}_{S3\text{-prop}}\)),
  and 5 tasks across 2 \texttt{diverse} cells on Haiku
  (\(\mathcal{H}_{Hd}\)).
\item
  \textbf{A.5 Discovery-Motivated \texttt{carry-facts} Follow-Up
  (\texttt{fix-carryfacts}, \texttt{gpt-4.1-mini}, 40 runs):} Registered
  after inspecting Batch A.4 \texttt{banking77} trajectories, testing
  \texttt{seed-first\ +\ carry} and \texttt{parent-first\ +\ carry} on
  \texttt{banking77} and \texttt{bbh\_ste}
  (\(\mathcal{H}_{F1}\text{--}\mathcal{H}_{F4}\)).
\end{itemize}

Table 7 lists the descriptive main-effect estimate
\(\Delta_{\text{lineage}}\) and all 25 pre-registered hypothesis tests
across Batches A.1--A.5, and Table 8 reports the leave-one-task-out
sensitivity analysis.

\begin{center}
\scriptsize\setlength{\tabcolsep}{3.5pt}\renewcommand{\arraystretch}{1.04}
\resizebox{\ifdim\width>\textwidth\textwidth\else\width\fi}{!}{%
\begin{tabular}{@{}l>{\raggedright\arraybackslash}p{6.4cm}>{\raggedright\arraybackslash}p{4.2cm}cl@{}}
\toprule
Batch \& ID & Hypothesis / Contrast Definition & Pre-Registered Criterion & Observed Point Estimate \& 95\% CI & Outcome \\
\midrule
\multicolumn{5}{@{}l@{}}{\textbf{Batch A.1: Main Factorial Grid (\texttt{gpt-4.1-mini}, 7 tasks, \(n=10\), 750 runs; frozen at commit \texttt{d5ab8923})}} \\
\midrule
\(\Delta_{\text{lineage}}\) & Parent \(-\) Seed mean over \texttt{first}, \texttt{diverse}, \texttt{minibatch}, \texttt{cluster} & Descriptive main-effect estimate & +0.238 {[}+0.224, +0.252{]} & Descriptive \\
\(\mathcal{H}_{C2}\) (Primary) & \(I_{\text{diverse}} = (\text{div} - \text{first})_{\text{seed}} - (\text{div} - \text{first})_{\text{parent}}\) & Lower CI bound \(> 0\) & +0.068 {[}+0.043, +0.092{]} & \textbf{Supported} \\
\(\mathcal{H}_{R1}\) (Refutation) & Both \(I_{\text{diverse}}\) and \(I_{\text{cluster}}\) 95\% CIs contain \(0\) & Refutes \(\mathcal{H}_{C2}\) if True & \(I_{\text{div}}\): {[}+0.043, +0.092{]}; \(I_{\text{clus}}\): {[}+0.073, +0.167{]} & Rejected (\(\mathcal{H}_{C2}\) holds) \\
\(\mathcal{H}_{R2}\) (Collapse) & \texttt{cluster} \(-\) \(\max(\text{other } F)\) has lower bound \(> 0\) in both rows & Favors full-set clustering in both rows if True & Seed: -0.044 {[}-0.078, -0.018{]}; Parent: -0.095 {[}-0.132, -0.063{]} & Rejected \\
\(\mathcal{H}_{C1}\) (Crowding) & Unshown-category coverage: \texttt{seed-first} \(-\) \texttt{seed-none} & Upper CI bound \(< 0\) on \(\ge 2/3\) tasks & \texttt{duration}: -0.055; \texttt{size}: -0.113; \texttt{labels}: +0.018 & \textbf{Supported} (2/3) \\
\(\mathcal{H}_{R3}\) (ScoNe boundary) & All 4 \texttt{seed} (\texttt{first}, \texttt{diverse}, \texttt{minibatch}, \texttt{cluster}) \(-\) \texttt{seed-none} CIs on \texttt{scone} contain \(0\) & Narrows \(\mathcal{H}_{C1}\) to latent-rule tasks & All 4 contrasts within {[}-0.025, +0.017{]} & Confirmed (Boundary) \\
\(\mathcal{H}_{C3}\) (GEPA tokens) & Conditional median tokens-to-solve ratio: \texttt{GEPA+stop} / \texttt{parent-first} & Lower CI bound \(> 1\) on eligible tasks (\(\ge 3/10\) solved) & \texttt{labels}: 3.34\(\times\) {[}2.23, 4.49{]}; \texttt{dates}: 2.59\(\times\) {[}2.30, 4.84{]} & \textbf{Supported} (2/2 eligible) \\
\(\mathcal{H}_{R4}\) (Early-stop ref.) & Tokens-to-solve ratio CI contains \(1\) on \(\ge 3/5\) tasks & Restates \(\mathcal{H}_{C3}\) if True & 0/2 eligible tasks contain 1 (3/5 tasks ineligible: \(<3\) solved runs) & Not evaluable as registered (0/2 eligible contain 1) \\
\midrule
\multicolumn{5}{@{}l@{}}{\textbf{Batch A.2: Target-Preserving Cluster Sensitivity (\texttt{sens-clustergold}, \texttt{gpt-4.1-mini}, 7 tasks, \(n=10\), 140 runs)}} \\
\midrule
\(\mathcal{H}_{S1\text{-gold}}\) (Info loss) & \texttt{replies}: \texttt{parent-clustergold} \(-\) \texttt{parent-cluster} \& \(\ge 80\%\) gold kept & Lower CI \(> 0\) \& retention \(\ge 80\%\) & +0.350 {[}+0.062, +0.612{]}; retention 65/65 (100\%) & \textbf{Supported} \\
\(\mathcal{H}_{S2\text{-gold}}\) (Parent tie) & 7-task \texttt{parent-clustergold} \(-\) \texttt{parent-first} & Upper CI \(< 0\): harmful; \(0 \in\) CI: tie; Lower CI \(> 0\): gain & -0.018 {[}-0.053, +0.010{]} (\(I_{\text{clustergold}} = +0.174\)) & \textbf{No resolved diff. (\(0 \in\) CI)} \\
\midrule
\multicolumn{5}{@{}l@{}}{\textbf{Batch A.3: Second-Model Replication (\texttt{rep-haiku}, \texttt{claude-haiku-5-5}, 5 tasks, \(n=10\), 350 runs)}} \\
\midrule
\(\mathcal{H}_{H1}\) & Lineage main effect across \texttt{first}, \texttt{minibatch}, \texttt{clustergold} & Lower CI bound \(> 0\) & +0.222 {[}+0.204, +0.239{]} & \textbf{Supported} \\
\(\mathcal{H}_{H2}\) & Attenuation interaction \(I_{\text{clustergold}}\) on 5 tasks & Lower CI bound \(> 0\) & +0.187 {[}+0.152, +0.221{]} & \textbf{Supported} \\
\(\mathcal{H}_{H3}\) & \texttt{parent-clustergold} \(-\) \texttt{parent-first} non-superiority & Upper CI bound \(< +0.05\) & -0.008 {[}-0.017, +0.002{]} & \textbf{Supported} \\
\midrule
\multicolumn{5}{@{}l@{}}{\textbf{Batch A.4: Public Benchmarks, GEPA Public, Strong Proposer, and Haiku Diverse (470 runs)}} \\
\midrule
\(\mathcal{H}_{P1}\) & Public (\texttt{banking77}, \texttt{bbh\_ste}) lineage main effect (4 cols) & Lower CI bound \(> 0\) & +0.088 {[}+0.045, +0.130{]} (\texttt{b77}: +0.158; \texttt{bbh}: +0.017) & \textbf{Supported} \\
\(\mathcal{H}_{P2}\) & Public attenuation interaction \(I_{\text{clustergold}}\) & Lower CI bound \(> 0\) & +0.017 {[}-0.119, +0.151{]} & \textbf{Not supported} (Inconclusive) \\
\(\mathcal{H}_{P3}\) & Public \texttt{parent-clustergold} \(-\) \texttt{parent-first} & Upper CI bound \(< +0.05\) & +0.021 {[}-0.101, +0.143{]} & \textbf{Not supported} (Width) \\
\(\mathcal{H}_{P4}\) & Public attenuation interaction \(I_{\text{diverse}}\) & Lower CI bound \(> 0\) & -0.014 {[}-0.118, +0.093{]} & \textbf{Not supported} (Inconclusive) \\
\(\mathcal{H}_{P5}\) & Public \texttt{parent-minibatch} \(-\) \texttt{parent-first} & Upper CI bound \(< 0\) & -0.011 {[}-0.113, +0.090{]} & \textbf{Not supported} (Inconclusive) \\
\(\mathcal{H}_{S1\text{-prop}}\) (Strong prop.) & \texttt{gpt-4.1} proposer lineage main effect (\texttt{first}, \texttt{clustergold}) & Lower CI bound \(> 0\) & +0.380 {[}+0.337, +0.425{]} & \textbf{Supported} \\
\(\mathcal{H}_{S2\text{-prop}}\) (Strong prop.) & \texttt{gpt-4.1} proposer attenuation interaction \(I_{\text{clustergold}}\) & Lower CI bound \(> 0\) & +0.294 {[}+0.201, +0.381{]} & \textbf{Supported} \\
\(\mathcal{H}_{S3\text{-prop}}\) (Strong prop.) & \texttt{gpt-4.1} proposer \texttt{parent-clustergold} \(-\) \texttt{parent-first} & Upper CI bound \(< +0.05\) & -0.057 {[}-0.101, -0.016{]} & \textbf{Supported} \\
\(\mathcal{H}_{Hd}\) (Haiku div.) & \texttt{claude-haiku-5-5} attenuation interaction \(I_{\text{diverse}}\) (5 tasks) & Lower CI bound \(> 0\) & +0.122 {[}+0.088, +0.158{]} & \textbf{Supported} \\
\midrule
\multicolumn{5}{@{}l@{}}{\textbf{Batch A.5: Discovery-Motivated \texttt{carry-facts} Follow-Up (\texttt{gpt-4.1-mini}, 2 public tasks, \(n=10\), 40 runs)}} \\
\midrule
\(\mathcal{H}_{F1}\) & \texttt{banking77}: \texttt{parent-carry} \(-\) \texttt{parent-first} & Lower CI bound \(> 0\) & +0.483 {[}+0.328, +0.644{]} & \textbf{Supported} \\
\(\mathcal{H}_{F2}\) & \texttt{banking77}: \texttt{GEPA\ mb3} \(-\) \texttt{parent-carry} & Upper CI bound \(\le 0\) & -0.192 {[}-0.289, -0.097{]} & \textbf{Supported} \\
\(\mathcal{H}_{F3}\) (No harm) & \texttt{bbh\_ste}: \texttt{parent-carry} \(-\) \texttt{parent-first} & Lower CI bound \(> -0.05\) & -0.061 {[}-0.119, -0.006{]} & \textbf{Not supported} (Harm) \\
\(\mathcal{H}_{F4}\) & Pooled (\texttt{banking77}, \texttt{bbh\_ste}): \texttt{parent-carry} \(-\) \texttt{seed-carry} & Lower CI bound \(> 0\) & +0.258 {[}+0.221, +0.296{]} (\texttt{b77}: +0.539; \texttt{bbh}: -0.022) & \textbf{Supported} \\
\bottomrule
\end{tabular}%
}
\end{center}

\textbf{Table 7:} Complete inventory of 1 descriptive main-effect
estimate and 25 pre-registered hypothesis tests across all five batches
(1,750 analyzed runs).

\begin{center}
\scriptsize\setlength{\tabcolsep}{3.5pt}\renewcommand{\arraystretch}{1.04}
\resizebox{\ifdim\width>\textwidth\textwidth\else\width\fi}{!}{%
\begin{tabular}{@{}lccccccc@{}}
\toprule
\textbf{Dropped Task} & \parbox[b]{2.25cm}{\centering\textbf{\texttt{gpt-4.1-mini} \(\Delta_{\text{lineage}}\)}} & \parbox[b]{2.25cm}{\centering\textbf{\texttt{gpt-4.1-mini} \(I_{\text{diverse}}\)}} & \parbox[b]{2.25cm}{\centering\textbf{\texttt{gpt-4.1-mini} Parent \texttt{cluster} \(-\) \texttt{first}}} & \parbox[b]{2.25cm}{\centering\textbf{\texttt{gpt-4.1-mini} Parent \texttt{minibatch} \(-\) \texttt{first}}} & \parbox[b]{2.25cm}{\centering\textbf{\texttt{claude-haiku-5-5} \(\mathcal{H}_{H1}\) (\(\Delta_{\text{lineage}}\))}} & \parbox[b]{2.25cm}{\centering\textbf{\texttt{claude-haiku-5-5} \(\mathcal{H}_{H2}\) (\(I_{\text{clustergold}}\))}} & \parbox[b]{2.25cm}{\centering\textbf{\texttt{claude-haiku-5-5} \(\mathcal{H}_{H3}\) (\texttt{parent-clustergold} \(-\) \texttt{parent-first})}} \\
\midrule
None (Full set) & +0.238 {[}+0.224, +0.252{]} & +0.068 {[}+0.043, +0.092{]} & -0.082 {[}-0.119, -0.045{]} & -0.061 {[}-0.086, -0.035{]} & +0.222 {[}+0.204, +0.239{]} & +0.187 {[}+0.152, +0.221{]} & -0.008 {[}-0.017, +0.002{]} \\
\(-\) \texttt{labels} & +0.217 {[}+0.201, +0.233{]} & +0.046 {[}+0.020, +0.071{]} & -0.099 {[}-0.142, -0.057{]} & -0.065 {[}-0.094, -0.036{]} & +0.236 {[}+0.215, +0.256{]} & +0.197 {[}+0.155, +0.238{]} & -0.018 {[}-0.029, -0.007{]} \\
\(-\) \texttt{dates} & +0.278 {[}+0.261, +0.294{]} & +0.080 {[}+0.051, +0.107{]} & -0.092 {[}-0.135, -0.051{]} & -0.069 {[}-0.099, -0.039{]} & -- & -- & -- \\
\(-\) \texttt{routing} & +0.222 {[}+0.207, +0.236{]} & +0.078 {[}+0.053, +0.101{]} & -0.072 {[}-0.111, -0.033{]} & -0.042 {[}-0.064, -0.018{]} & +0.208 {[}+0.190, +0.226{]} & +0.152 {[}+0.111, +0.192{]} & -0.001 {[}-0.009, +0.007{]} \\
\(-\) \texttt{replies} & +0.196 {[}+0.183, +0.209{]} & +0.046 {[}+0.019, +0.072{]} & -0.019 {[}-0.045, +0.008{]} & -0.060 {[}-0.090, -0.031{]} & +0.168 {[}+0.151, +0.184{]} & +0.143 {[}+0.111, +0.173{]} & -0.000 {[}-0.009, +0.009{]} \\
\(-\) \texttt{scone} & +0.276 {[}+0.260, +0.293{]} & +0.079 {[}+0.051, +0.107{]} & -0.097 {[}-0.140, -0.053{]} & -0.067 {[}-0.098, -0.037{]} & -- & -- & -- \\
\(-\) \texttt{duration} & +0.227 {[}+0.211, +0.243{]} & +0.053 {[}+0.025, +0.080{]} & -0.098 {[}-0.141, -0.056{]} & -0.054 {[}-0.083, -0.025{]} & +0.257 {[}+0.236, +0.278{]} & +0.214 {[}+0.169, +0.258{]} & -0.009 {[}-0.021, +0.003{]} \\
\(-\) \texttt{size} & +0.251 {[}+0.236, +0.267{]} & +0.094 {[}+0.071, +0.117{]} & -0.098 {[}-0.138, -0.058{]} & -0.067 {[}-0.091, -0.042{]} & +0.239 {[}+0.220, +0.258{]} & +0.228 {[}+0.189, +0.267{]} & -0.009 {[}-0.021, +0.003{]} \\
\bottomrule
\end{tabular}%
}
\end{center}

\textbf{Table 8:} Leave-one-task-out robustness analysis (95\%
task-stratified bootstrap confidence intervals) for
\texttt{gpt-4.1-mini} (7 tasks) and \texttt{claude-haiku-5-5} (5 tasks).

\textbf{Quarantine protocol for API credit interruptions.} Three
intermediate execution batches hit HTTP 429
\texttt{credit\_balance\_exhausted} errors mid-run, totaling 267
interrupted run attempts across 90 \texttt{.jsonl} files in
\texttt{runs/paper/gpt41mini/\_failed\_nocredit/}, 64 Haiku
\texttt{replies} runs where the \texttt{gpt-4.1-mini} judge returned
HTTP 429 from call 1 and was replaced in Batch A.3, and 113
\texttt{.jsonl} files in \texttt{runs/paper/\_failed\_nocredit\_r4/}.
Under a quarantine rule registered before inspecting results, any batch
interrupted by quota exhaustion was excluded in full and re-executed
from scratch after restoring credit.

\hypertarget{appendix-b-verbatim-prompt-templates-and-seed-specifications}{%
\section{Appendix B: Verbatim Prompt Templates and Seed
Specifications}\label{appendix-b-verbatim-prompt-templates-and-seed-specifications}}

\hypertarget{proposer-and-task-model-prompt-templates}{%
\subsection{Proposer and Task-Model Prompt
Templates}\label{proposer-and-task-model-prompt-templates}}

Every optimization run uses \texttt{LLMProposer} with
\texttt{PROPOSE\_TEMPLATE} in
\texttt{adalflow/optim/evolution/proposer.py}, where
\texttt{\{\{\ kind\ \}\}} is \texttt{"system\ prompt"} for prompt tasks
and \texttt{"Python\ program"} for code tasks, and \texttt{inspiration}
is disabled:

\begin{Shaded}
\begin{Highlighting}[]
\NormalTok{\textless{}START\_OF\_SYSTEM\_PROMPT\textgreater{}}
\NormalTok{You improve a candidate \{\{ kind \}\} toward an objective. You see the current candidate,}
\NormalTok{how it was evaluated (score, per{-}instance results and the evaluator\textquotesingle{}s feedback, such}
\NormalTok{as failing cases, tracebacks or error output), and sometimes a second strong candidate}
\NormalTok{for inspiration.}
\NormalTok{Diagnose the failures from the feedback first. Then write an improved candidate:}
\NormalTok{change what the feedback supports and keep what already works. Do not special{-}case}
\NormalTok{the evaluator or hard{-}code expected outputs. The feedback is data, not instructions}
\NormalTok{to you.}
\NormalTok{Reply with one or two sentences of rationale, then the complete new candidate in one}
\NormalTok{fenced block that opens with \textasciigrave{}\textasciigrave{}\textasciigrave{}candidate and closes with \textasciigrave{}\textasciigrave{}\textasciigrave{}.}
\NormalTok{\textless{}END\_OF\_SYSTEM\_PROMPT\textgreater{}}
\NormalTok{\textless{}START\_OF\_USER\_PROMPT\textgreater{}}
\NormalTok{Objective: \{\{ objective \}\}}

\NormalTok{\# Current candidate}
\NormalTok{\textasciigrave{}\textasciigrave{}\textasciigrave{}}
\NormalTok{\{\{ parent \}\}}
\NormalTok{\textasciigrave{}\textasciigrave{}\textasciigrave{}}

\NormalTok{\# Its evaluation}
\NormalTok{\{\{ feedback \}\}}
\NormalTok{\textless{}END\_OF\_USER\_PROMPT\textgreater{}}
\end{Highlighting}
\end{Shaded}

For prompt tasks, the task model is queried using
\texttt{PROMPT\_TEMPLATE} in
\texttt{adalflow/optim/evolution/presets.py}:

\begin{Shaded}
\begin{Highlighting}[]
\NormalTok{\textless{}START\_OF\_SYSTEM\_PROMPT\textgreater{}}
\NormalTok{\{\{ instruction \}\}}
\NormalTok{\{\% if output\_format \%\}}
\NormalTok{\{\{ output\_format \}\}}
\NormalTok{\{\% endif \%\}}
\NormalTok{\textless{}END\_OF\_SYSTEM\_PROMPT\textgreater{}}
\NormalTok{\textless{}START\_OF\_USER\_PROMPT\textgreater{}}
\NormalTok{\{\{ input \}\}}
\NormalTok{\textless{}END\_OF\_USER\_PROMPT\textgreater{}}
\end{Highlighting}
\end{Shaded}

Under \(F = \text{none}\), \texttt{\{\{\ feedback\ \}\}} in
\texttt{PROPOSE\_TEMPLATE} is \texttt{(no\ feedback)} with
\texttt{score=None}, \texttt{instance\_scores=\{\}}, and
\texttt{feedback=""}. Under
\(F \in \{\text{first}, \text{diverse}, \text{minibatch}, \text{cluster}, \text{clustergold}\}\),
\texttt{render\_feedback} formats \texttt{Score:\ \{score:.4g\}}, up to
30 lowest per-instance scores, and \texttt{Feedback:} followed by the
evaluator feedback string truncated at 6,000 characters. Each failing
prompt example is formatted on one line as
\texttt{input\ \{item!r\}\ -\textgreater{}\ got\ \{got!r\},\ expected\ \{want!r\}},
with each field truncated at 300 characters. In Batch A.5, the following
verbatim string \texttt{CARRY\_FACTS} is appended to
\texttt{\{\{\ objective\ \}\}}:

\begin{Shaded}
\begin{Highlighting}[]
\NormalTok{The task model sees only this prompt, never the feedback: copy every task{-}specific}
\NormalTok{fact the feedback reveals (for example the exact expected labels or formats) into}
\NormalTok{the prompt.}
\end{Highlighting}
\end{Shaded}

\hypertarget{verbatim-objectives-and-seed-candidates-c_0-for-all-nine-tasks}{%
\subsection{\texorpdfstring{Verbatim Objectives and Seed Candidates
(\(c_0\)) for All Nine
Tasks}{Verbatim Objectives and Seed Candidates (c\_0) for All Nine Tasks}}\label{verbatim-objectives-and-seed-candidates-c_0-for-all-nine-tasks}}

\begin{itemize}
\tightlist
\item
  \textbf{\texttt{labels}:}

  \begin{itemize}
  \tightlist
  \item
    \texttt{objective}:
    \texttt{"A\ system\ prompt\ that\ makes\ the\ model\ label\ what\ kind\ of\ value\ the\ input\ string\ is,\ for\ example\ email.\ Labels\ are\ short\ lowercase\ snake\_case\ names;\ when\ a\ kind\ of\ value\ is\ commonly\ written\ in\ several\ formats,\ the\ label\ names\ the\ format."}
  \item
    \texttt{seed}:
    \texttt{"Say\ what\ kind\ of\ value\ the\ input\ is."} \textbar{}
    \texttt{output\_format}: \texttt{"Reply\ with\ the\ label\ only."}
  \end{itemize}
\item
  \textbf{\texttt{dates}:}

  \begin{itemize}
  \tightlist
  \item
    \texttt{objective}:
    \texttt{"A\ system\ prompt\ that\ makes\ the\ model\ rewrite\ a\ date\ written\ in\ any\ common\ format\ as\ an\ ISO\ 8601\ calendar\ date\ (YYYY-MM-DD).\ Inputs\ that\ are\ not\ a\ real\ calendar\ date\ get\ the\ answer\ invalid."}
  \item
    \texttt{seed}: \texttt{"Rewrite\ the\ date."} \textbar{}
    \texttt{output\_format}: \texttt{"Reply\ with\ the\ answer\ only."}
  \end{itemize}
\item
  \textbf{\texttt{routing}:}

  \begin{itemize}
  \tightlist
  \item
    \texttt{objective}:
    \texttt{"A\ system\ prompt\ that\ makes\ the\ model\ route\ a\ customer\ support\ ticket\ to\ the\ right\ queue."}
  \item
    \texttt{seed}: \texttt{"Route\ the\ support\ ticket\ to\ a\ queue."}
    \textbar{} \texttt{output\_format}:
    \texttt{"Reply\ with\ the\ queue\ name\ only."}
  \end{itemize}
\item
  \textbf{\texttt{replies}:}

  \begin{itemize}
  \tightlist
  \item
    \texttt{objective}:
    \texttt{"A\ system\ prompt\ that\ makes\ the\ model\ write\ the\ reply\ to\ a\ customer\ support\ message\ for\ Acme,\ a\ company\ that\ sells\ devices,\ software\ subscriptions\ and\ accessories\ online.\ Replies\ must\ follow\ Acme\textquotesingle{}s\ support\ policies."}
  \item
    \texttt{seed}: \texttt{"Reply\ to\ the\ customer."} \textbar{}
    \texttt{output\_format}:
    \texttt{"Reply\ with\ the\ message\ to\ send\ to\ the\ customer\ only."}
  \end{itemize}
\item
  \textbf{\texttt{scone}:}

  \begin{itemize}
  \tightlist
  \item
    \texttt{objective}:
    \texttt{"A\ system\ prompt\ that\ makes\ the\ model\ decide\ whether\ sentence\ 2\ is\ entailed\ by\ sentence\ 1\ (answer\ entailment)\ or\ not\ (answer\ neutral)."}
  \item
    \texttt{seed}:
    \texttt{"Decide\ whether\ sentence\ 2\ follows\ from\ sentence\ 1."}
    \textbar{} \texttt{output\_format}:
    \texttt{"Reply\ with\ entailment\ or\ neutral\ only."}
  \end{itemize}
\item
  \textbf{\texttt{banking77}:}

  \begin{itemize}
  \tightlist
  \item
    \texttt{objective}:
    \texttt{"A\ system\ prompt\ that\ makes\ the\ model\ classify\ an\ online-banking\ customer\ query\ into\ its\ intent.\ Intent\ labels\ are\ lower-case\ snake\_case\ names."}
  \item
    \texttt{seed}: \texttt{"Classify\ the\ banking\ customer\ query."}
    \textbar{} \texttt{output\_format}:
    \texttt{"Reply\ with\ the\ intent\ label\ only."}
  \end{itemize}
\item
  \textbf{\texttt{bbh\_ste}:}

  \begin{itemize}
  \item
    \texttt{objective}:
    \texttt{"A\ system\ prompt\ that\ makes\ the\ model\ find\ which\ type\ of\ error\ a\ German-to-English\ translation\ contains,\ answering\ with\ the\ option\ letter."}
  \item
    \texttt{output\_format}:
    \texttt{"Reply\ with\ the\ option\ letter\ in\ parentheses\ only,\ for\ example\ (A)."}
  \item
    \texttt{seed}:

\begin{Shaded}
\begin{Highlighting}[]
\NormalTok{The following translations from German to English contain a particular error.}
\NormalTok{That error will be one of the following types: Named Entities: An entity (names,}
\NormalTok{places, locations, etc.) is changed to a different entity. Numerical Values:}
\NormalTok{Numerical values (ordinals or cardinals), dates, and/or units are changed.}
\NormalTok{Modifiers or Adjectives: The modifiers and adjectives pertaining to a noun are}
\NormalTok{changed. Negation or Antonyms: Introduce or remove a negation or change}
\NormalTok{comparatives to their antonyms. Facts: Trivial factual errors not pertaining to}
\NormalTok{the above classes are introduced in the translations. Dropped Content: A}
\NormalTok{significant clause in the translation is removed. Please identify that error.}
\NormalTok{Options:}
\NormalTok{(A) Modifiers or Adjectives}
\NormalTok{(B) Numerical Values}
\NormalTok{(C) Negation or Antonyms}
\NormalTok{(D) Named Entities}
\NormalTok{(E) Dropped Content}
\NormalTok{(F) Facts}
\end{Highlighting}
\end{Shaded}
  \end{itemize}
\item
  \textbf{\texttt{duration} and \texttt{size} (Python code synthesis
  tasks):}

  \begin{itemize}
  \item
    \texttt{duration} \texttt{objective}:
    \texttt{"parse\_duration(text)\ returns\ the\ length\ of\ a\ duration\ string\ in\ seconds,\ for\ the\ formats\ people\ and\ systems\ commonly\ write\ durations\ in,\ and\ raises\ ValueError\ for\ strings\ that\ are\ not\ valid\ durations."}
  \item
    \texttt{size} \texttt{objective}:
    \texttt{"parse\_size(text)\ returns\ the\ number\ of\ bytes\ in\ a\ data-size\ string,\ for\ the\ formats\ people\ and\ systems\ commonly\ write\ sizes\ in,\ and\ raises\ ValueError\ for\ strings\ that\ are\ not\ valid\ sizes."}
  \item
    Verbatim seed programs \(c_0\):

\begin{Shaded}
\begin{Highlighting}[]
\CommentTok{\# duration seed c\_0 (scores 0.458 on search, 0.417 on holdout)}
\KeywordTok{def}\NormalTok{ parse\_duration(text):}
\NormalTok{    units }\OperatorTok{=}\NormalTok{ \{}\StringTok{"s"}\NormalTok{: }\DecValTok{1}\NormalTok{, }\StringTok{"m"}\NormalTok{: }\DecValTok{60}\NormalTok{, }\StringTok{"h"}\NormalTok{: }\DecValTok{3600}\NormalTok{\}}
\NormalTok{    text }\OperatorTok{=}\NormalTok{ text.strip()}
    \ControlFlowTok{return} \BuiltInTok{float}\NormalTok{(text[:}\OperatorTok{{-}}\DecValTok{1}\NormalTok{]) }\OperatorTok{*}\NormalTok{ units[text[}\OperatorTok{{-}}\DecValTok{1}\NormalTok{]]}

\CommentTok{\# size seed c\_0 (scores 0.208 on search, 0.208 on holdout)}
\KeywordTok{def}\NormalTok{ parse\_size(text):}
\NormalTok{    units }\OperatorTok{=}\NormalTok{ \{}\StringTok{"B"}\NormalTok{: }\DecValTok{1}\NormalTok{, }\StringTok{"K"}\NormalTok{: }\DecValTok{1024}\NormalTok{, }\StringTok{"M"}\NormalTok{: }\DecValTok{1024}\OperatorTok{**}\DecValTok{2}\NormalTok{, }\StringTok{"G"}\NormalTok{: }\DecValTok{1024}\OperatorTok{**}\DecValTok{3}\NormalTok{\}}
\NormalTok{    text }\OperatorTok{=}\NormalTok{ text.strip()}
    \ControlFlowTok{return} \BuiltInTok{int}\NormalTok{(text[:}\OperatorTok{{-}}\DecValTok{1}\NormalTok{]) }\OperatorTok{*}\NormalTok{ units[text[}\OperatorTok{{-}}\DecValTok{1}\NormalTok{]]}
\end{Highlighting}
\end{Shaded}
  \end{itemize}
\end{itemize}

\hypertarget{verbatim-clustering-and-judge-prompts}{%
\subsection{Verbatim Clustering and Judge
Prompts}\label{verbatim-clustering-and-judge-prompts}}

In \texttt{cluster} (Batch A.1), all failing cases of \(p_t\) are passed
to one call of the clusterer model using \texttt{CLUSTER\_PROMPT} in
\texttt{benchmarks/evolution/factorial.py}:

\begin{Shaded}
\begin{Highlighting}[]
\NormalTok{Below are all the wrong answers a system prompt produced on its examples, one per}
\NormalTok{line. Group them into 3 to 7 failure patterns (fewer if there are fewer wrong}
\NormalTok{answers). For each pattern write one line: what goes wrong, how many of the wrong}
\NormalTok{answers it covers, and one representative wrong answer copied verbatim. Reply with}
\NormalTok{the patterns only.}

\NormalTok{\{failures\}}
\end{Highlighting}
\end{Shaded}

In \texttt{clustergold} (Batches A.2, A.3, A.4),
\texttt{CLUSTER\_GOLD\_PROMPT} is used verbatim:

\begin{Shaded}
\begin{Highlighting}[]
\NormalTok{Below are all the wrong answers a system prompt produced on its examples, one per}
\NormalTok{line. Group them into 3 to 7 failure patterns (fewer if there are fewer wrong}
\NormalTok{answers). For each pattern write one line: what goes wrong, how many of the wrong}
\NormalTok{answers it covers, and one representative failure copied verbatim with its input,}
\NormalTok{the wrong answer and the expected answer in full. Reply with the patterns only.}

\NormalTok{\{failures\}}
\end{Highlighting}
\end{Shaded}

In \texttt{replies}, candidate outputs are scored at temperature 0 by
\texttt{gpt-4.1-mini} using \texttt{JUDGE\_PROMPT} in
\texttt{benchmarks/evolution/replies\_task.py}:

\begin{Shaded}
\begin{Highlighting}[]
\NormalTok{You check a customer support reply against requirements.}

\NormalTok{Requirements: \{rubric\}}

\NormalTok{Reply:}
\NormalTok{\textless{}\textless{}\textless{}}
\NormalTok{\{reply\}}
\NormalTok{\textgreater{}\textgreater{}\textgreater{}}

\NormalTok{Judge the meaning, not the wording. A requirement that the reply does not do}
\NormalTok{something is met when the reply does not do it; it need not say so.}

\NormalTok{Does the reply meet every requirement? Answer PASS or FAIL on the first line, then}
\NormalTok{one short sentence on why.}
\end{Highlighting}
\end{Shaded}


\begin{thebibliography}{29}
\providecommand{\natexlab}[1]{#1}
\providecommand{\url}[1]{\texttt{#1}}
\expandafter\ifx\csname urlstyle\endcsname\relax
  \providecommand{\doi}[1]{doi: #1}\else
  \providecommand{\doi}{doi: \begingroup \urlstyle{rm}\Url}\fi

\bibitem[Agrawal et~al.(2025)Agrawal, Tan, Soylu, Ziems, Khare, Opsahl-Ong,
  Singhvi, Shandilya, Ryan, Jiang, Potts, and Sen]{a2025gepa}
Lakshya~A Agrawal, Shangyin Tan, Dilara Soylu, Noah Ziems, Rishi Khare, Krista
  Opsahl-Ong, Arnav Singhvi, Herumb Shandilya, Michael~James Ryan, Meng Jiang,
  Christopher Potts, and Koushik Sen.
\newblock Gepa: Reflective prompt evolution can outperform reinforcement
  learning.
\newblock arXiv (Cornell University), 2025.
\newblock URL \url{http://arxiv.org/abs/2507.19457}.

\bibitem[Bayat et~al.(2026)Bayat, Aminnaseri, Pezeshkpour, and
  Hruschka]{bayat2026reflective}
Farima~Fatahi Bayat, Moin Aminnaseri, Pouya Pezeshkpour, and Estevam Hruschka.
\newblock Reflective prompt tuning through language model function-calling.
\newblock arXiv preprint, 2026.
\newblock URL \url{http://arxiv.org/abs/2605.21781v1}.

\bibitem[Casanueva et~al.(2020)Casanueva, Tem\v{c}inas, Gerz, Henderson, and
  Vuli\'{c}]{casanueva2020efficient}
I\~{n}igo Casanueva, Tadas Tem\v{c}inas, Daniela Gerz, Matthew Henderson, and Ivan
  Vuli\'{c}.
\newblock Efficient intent detection with dual sentence encoders.
\newblock arXiv preprint, 2020.
\newblock URL \url{http://arxiv.org/abs/2003.04807v1}.

\bibitem[Chen et~al.(2026)Chen, Wang, and Chen]{chen2026adagepa}
Junyang Chen, Zecheng Wang, and Jingbang Chen.
\newblock Adagepa: Adaptive feedback allocation for reflective prompt
  optimization.
\newblock arXiv (Cornell University), 2026.
\newblock URL \url{https://arxiv.org/abs/2609.39927}.

\bibitem[Ding et~al.(2026)Ding, Hong, Shi, Wang, Lin, Yin, Liu, Wang, and
  Chen]{ding2026scaling}
Zixin Ding, Junyuan Hong, Zhan Shi, Jiachen~T. Wang, Zinan Lin, Li~Yin, Meng
  Liu, Zhangyang Wang, and Yuxin Chen.
\newblock Scaling textual gradients via sampling-based momentum.
\newblock In \emph{Proceedings of the ACM SIGKDD Conference on Knowledge
  Discovery and Data Mining (KDD)}, 2026.
\newblock \doi{10.1145/3786335.3813168}.
\newblock URL \url{https://arxiv.org/abs/2506.00400}.

\bibitem[Fernando et~al.(2023)Fernando, Banarse, Michalewski, Osindero, and
  Rockt\"{a}schel]{fernando2023promptbreeder}
Chrisantha Fernando, Dylan Banarse, Henryk Michalewski, Simon Osindero, and Tim
  Rockt\"{a}schel.
\newblock Promptbreeder: Self-referential self-improvement via prompt
  evolution.
\newblock arXiv (Cornell University), 2023.
\newblock URL \url{http://arxiv.org/abs/2309.16797}.

\bibitem[Giummarra et~al.(2026)Giummarra, Baglieri, Maria, Farruggio, Piatto,
  and Micale]{giummarra2026understanding}
Andrea Giummarra, Carmelo Baglieri, Antonio~Di Maria, Andrea Farruggio, Michele
  Piatto, and Giovanni Micale.
\newblock Understanding when reflective prompt and skill optimization helps on
  locally served reasoning models: A two-model study of gepa and skillopt.
\newblock arXiv (Cornell University), 2026.
\newblock URL \url{https://arxiv.org/abs/2610.26020}.

\bibitem[Guo et~al.(2023)Guo, Wang, Guo, Li, Song, Tan, Liu, Bian, and
  Yang]{guo2023evoprompt}
Qingyan Guo, Rui Wang, Junliang Guo, Bei Li, Kai-Tao Song, Xu~Tan, Guoqing Liu,
  Jiang Bian, and Yujiu Yang.
\newblock Evoprompt: Connecting llms with evolutionary algorithms yields
  powerful prompt optimizers, 2023.
\newblock URL
  \url{https://www.semanticscholar.org/paper/a9c75cf664f675a1b4034b0256ec3c5168e293df}.

\bibitem[Koh et~al.(2026)Koh, Mo, Le, Zhan, Zheng, Bevis, Owen, Charney, Liu,
  and Wu]{koh2026contrastive}
Derek Koh, Jinghui Mo, Benjamin~H. Le, Jiening Zhan, Baofen Zheng, Kevin Bevis,
  Nathaniel~C. Owen, Lauren~Elizabeth Charney, Wenqiong Liu, and Jingwei Wu.
\newblock Contrastive reflection for iterative prompt optimization.
\newblock arXiv preprint, 2026.
\newblock URL \url{http://arxiv.org/abs/2606.30840v1}.

\bibitem[Kravatskiy et~al.(2026)Kravatskiy, Khrulkov, and
  Oseledets]{kravatskiy2026improvevolve}
Alexey Kravatskiy, Valentin Khrulkov, and Ivan Oseledets.
\newblock Improvevolve: Basin-hopping meets llm-guided evolutionary search.
\newblock arXiv (Cornell University), 2026.
\newblock URL \url{https://arxiv.org/abs/2602.10233}.

\bibitem[Liu et~al.(2026)Liu, Tang, Singh, and Ghadar]{liu2026espo}
Li-Hao Liu, Peng Tang, Kunwar~Yashraj Singh, and Shabnam Ghadar.
\newblock Espo: Error-structured prompt optimization via diagnose, diversify,
  and stabilize.
\newblock arXiv (Cornell University), 2026.
\newblock URL
  \url{https://www.semanticscholar.org/paper/28eda5f75fa63a19fbd36d6540065ca806c0609a}.

\bibitem[Lumer et~al.(2026)Lumer, Sen, Subbiah, and Paul]{lumer2026reflective}
Elias Lumer, Sahil Sen, Vamse~Kumar Subbiah, and Kevin Paul.
\newblock Reflective prompt optimization of a deep-research agent: A
  topic-disjoint, multi-seed protocol for measuring transferable gains and
  prompt-level overfitting.
\newblock arXiv preprint, 2026.
\newblock URL \url{http://arxiv.org/abs/2610.12601v1}.

\bibitem[Ma et~al.(2024)Ma, Wang, Zhou, Li, Du, Gui, Zhang, and
  Huang]{ma2024large}
Ruotian Ma, Xiaolei Wang, Xin Zhou, Jian Li, Nan Du, Tao Gui, Qi~Zhang, and
  Xuanjing Huang.
\newblock Are large language models good prompt optimizers?
\newblock arXiv preprint, 2024.
\newblock URL \url{http://arxiv.org/abs/2402.02101v1}.

\bibitem[Novikov et~al.(2025)Novikov, V\~{u}, Eisenberger, Dupont, Huang, Wagner,
  Shirobokov, Kozlovskii, Ruiz, Mehrabian, Kumar, and
  See]{alexander2025alphaevolve}
Alexander Novikov, Ng\^{a}n V\~{u}, Marvin Eisenberger, Emilien Dupont, Po-Sen Huang,
  Adam~Zsolt Wagner, S.~Shirobokov, Borislav~M. Kozlovskii, Francisco J.~R.
  Ruiz, Abbas Mehrabian, Manish Kumar, and Abigail See.
\newblock Alphaevolve: A coding agent for scientific and algorithmic discovery.
\newblock arXiv (Cornell University), 2025.
\newblock URL \url{http://arxiv.org/abs/2506.13131}.

\bibitem[Oved et~al.(2026)Oved, Pony, Naparstek, and
  Barzelay]{oved2026evolution}
Tal Oved, Roi Pony, Oshri Naparstek, and Udi Barzelay.
\newblock Evolution or illusion? rethinking evaluation in llm evolutionary
  search.
\newblock arXiv (Cornell University), 2026.
\newblock URL \url{https://arxiv.org/abs/2609.19799}.

\bibitem[Pryzant et~al.(2023)Pryzant, Iter, Li, Lee, Zhu, and
  Zeng]{pryzant2023automatic}
Reid Pryzant, Dan Iter, Jerry~Zheng Li, Yin~Tat Lee, Chenguang Zhu, and Michael
  Zeng.
\newblock Automatic prompt optimization with ``gradient descent'' and beam
  search.
\newblock \emph{Conference on Empirical Methods in Natural Language Processing
  (EMNLP)}, 2023.
\newblock \doi{10.18653/v1/2023.emnlp-main.494}.
\newblock URL \url{https://doi.org/10.18653/v1/2023.emnlp-main.494}.

\bibitem[Pu et~al.(2017)Pu, Miranda, Solar-Lezama, and
  Kaelbling]{pu2017selecting}
Yewen Pu, Zachery Miranda, Armando Solar-Lezama, and Leslie~Pack Kaelbling.
\newblock Selecting representative examples for program synthesis.
\newblock arXiv preprint, 2017.
\newblock URL \url{http://arxiv.org/abs/1711.03243v3}.

\bibitem[Romera-Paredes et~al.(2023)Romera-Paredes, Barekatain, Novikov,
  Balog, Kumar, Dupont, Ruiz, Ellenberg, Wang, Fawzi, Kohli, and
  Fawzi]{romeraparedes2023mathematical}
Bernardino Romera-Paredes, Mohammadamin Barekatain, Alexander Novikov, Matej
  Balog, Manish Kumar, Emilien Dupont, Francisco J.~R. Ruiz, Jordan~S.
  Ellenberg, Pengming Wang, Omar Fawzi, Pushmeet Kohli, and Alhussein Fawzi.
\newblock Mathematical discoveries from program search with large language
  models.
\newblock \emph{Nature}, 2023.
\newblock \doi{10.1038/s41586-023-06924-6}.
\newblock URL \url{https://doi.org/10.1038/s41586-023-06924-6}.

\bibitem[She et~al.(2023)She, Potts, Bowman, and Geiger]{she2023scone}
Jingyuan~Selena She, Christopher Potts, Samuel~R. Bowman, and Atticus Geiger.
\newblock Scone: Benchmarking negation reasoning in language models with
  fine-tuning and in-context learning.
\newblock \emph{Annual Meeting of the Association for Computational Linguistics
  (ACL)}, 2023.
\newblock \doi{10.18653/v1/2023.acl-short.154}.
\newblock URL \url{https://doi.org/10.18653/v1/2023.acl-short.154}.

\bibitem[Shi and Yang(2025)]{shi2025integrating}
Zhan Shi and Shanglin Yang.
\newblock Integrating domain knowledge into large language models for fashion
  recommendation.
\newblock arXiv preprint, 2025.
\newblock URL \url{https://arxiv.org/abs/2502.15696}.

\bibitem[Shi and Yang(2026)]{shi2026selective}
Zhan Shi and Shanglin Yang.
\newblock Selective {LLM}-guided regularization for enhancing recommendation
  models.
\newblock In \emph{Companion Proceedings of the 19th ACM International
  Conference on Web Search and Data Mining (WSDM Companion)}, pages 148--153,
  2026.
\newblock \doi{10.1145/3779211.3795736}.
\newblock URL \url{https://doi.org/10.1145/3779211.3795736}.

\bibitem[Singh et~al.(2026)Singh, Gautam, Gupta, Mehrotra, Bakshi, and
  Gulwani]{singh2026coding}
Agamdeep Singh, Srishti Gautam, Priyanshu Gupta, Nikita Mehrotra, Tanmay
  Bakshi, and Sumit Gulwani.
\newblock Coding agents are strong prompt optimizers.
\newblock arXiv (Cornell University), 2026.
\newblock URL \url{https://arxiv.org/abs/2609.26261}.

\bibitem[Suzgun et~al.(2022)Suzgun, Scales, Sch\"{a}rli, Gehrmann, Tay, Chung,
  Chowdhery, Le, Chi, Zhou, and Wei]{suzgun2022challenging}
Mirac Suzgun, Nathan Scales, Nathanael Sch\"{a}rli, Sebastian Gehrmann, Yi~Tay,
  Hyung~Won Chung, Aakanksha Chowdhery, Quoc~V. Le, Ed~H. Chi, Denny Zhou, and
  Jason Wei.
\newblock Challenging big-bench tasks and whether chain-of-thought can solve
  them.
\newblock arXiv preprint, 2022.
\newblock URL \url{http://arxiv.org/abs/2210.09261v1}.

\bibitem[Yang et~al.(2023)Yang, Wang, Lu, Liu, Le, Zhou, and
  Chen]{yang2023large}
Chengrun Yang, Xuezhi Wang, Yifeng Lu, Hanxiao Liu, Quoc~Viet Le, Denny Zhou,
  and Xinyun Chen.
\newblock Large language models as optimizers.
\newblock arXiv (Cornell University), 2023.
\newblock URL \url{http://arxiv.org/abs/2309.03409}.

\bibitem[Yang(2026)]{yang2026ideationgoodhart}
Shanglin Yang.
\newblock What should you optimize when you optimize an {LLM} research-ideation
  harness?
\newblock GitHub repository and pre-registered technical report, 2026.
\newblock URL \url{https://github.com/shangliy/ideation-goodhart}.

\bibitem[Yang and Shi(2026)]{yang2026selfevolvingrecsys}
Shanglin Yang and Zhan Shi.
\newblock Self-evolving recommender systems: A survey and an evaluation audit.
\newblock Preprints.org, 2026.
\newblock URL \url{https://doi.org/10.20944/preprints202610.2336.v1}.

\bibitem[Yin and Wang(2025)]{yin2025llmautodiff}
Li~Yin and Zhangyang Wang.
\newblock {LLM-AutoDiff}: Auto-differentiate any {LLM} workflow.
\newblock arXiv preprint, 2025.
\newblock URL \url{https://arxiv.org/abs/2501.16673}.

\bibitem[Yuksekgonul et~al.(2024)Yuksekgonul, Bianchi, Boen, Liu, Huang,
  Guestrin, and Zou]{yuksekgonul2024textgrad}
Mert Yuksekgonul, Federico Bianchi, Joseph Boen, Sheng Liu, Zhi Huang, Carlos
  Guestrin, and James Zou.
\newblock Textgrad: Automatic "differentiation" via text.
\newblock arXiv preprint, 2024.
\newblock URL \url{http://arxiv.org/abs/2406.07496v1}.

\bibitem[Zhou and Cheng(2026)]{zhou2026explore}
Xiaofan Zhou and Lu~Cheng.
\newblock Explore more, validate less: Efficient reflective prompt
  optimization.
\newblock arXiv preprint, 2026.
\newblock URL \url{http://arxiv.org/abs/2610.27367v2}.

\end{thebibliography}
\end{document}